\documentclass[a4paper,12pt,fleqn]{article}
\usepackage[margin=1.25in]{geometry}

\usepackage{setspace}
\usepackage{lineno}

\usepackage[utf8]{inputenc}
\usepackage[USenglish]{babel}

\usepackage{amssymb}
\usepackage{amsfonts}
\usepackage{amsmath}
\usepackage{bbm}
\usepackage{comment}
\usepackage{graphicx}
\usepackage{tgpagella}
\usepackage{titlesec}
\usepackage{amsthm}
\usepackage{mathrsfs}  
\usepackage{enumitem}
\usepackage{mathabx}

\usepackage[authoryear]{natbib}
\usepackage[colorlinks=true,citecolor=maastricht,linkcolor=darkscarlet,pagebackref=true]{hyperref}
\usepackage[nameinlink]{cleveref}

\usepackage{thmtools}

\renewcommand\thmcontinues[1]{Continued}

\usepackage{xcolor}
\definecolor{darkred}{rgb}{0.55, 0.0, 0.0}

\definecolor{darkscarlet}{rgb}{0.34, 0.01, 0.1}

\let\OLDthebibliography\thebibliography
\renewcommand\thebibliography[1]{
  \OLDthebibliography{#1}
  \setlength{\parskip}{0pt}
  \setlength{\itemsep}{0pt plus 0.5ex}
}

\usepackage[]{nameref}
\newcommand*{\fullref}[1]{\hyperref[{#1}]{\autoref*{#1} \nameref*{#1}}}

\usepackage{titletoc}
\definecolor{maastricht}{rgb}{0.0, 0.18, 0.39}
\theoremstyle{plain}
\newtheorem{theorem}{Theorem}

\newtheorem{corollary}{Corollary}

\newtheorem{lemma}{Lemma}

\theoremstyle{definition}
\newtheorem{definition}{Definition}
\newtheorem{example}{Example}

\usepackage{tikz}
\usepackage{pstricks,egameps}
\usepackage{pst-node,pst-tree}
\usetikzlibrary{calc}
\usetikzlibrary{matrix}
\usetikzlibrary{positioning}
\usetikzlibrary{arrows}
\usepackage{epigraph}
\usepackage{enumitem}
\usepackage{tcolorbox}

\begin{document}

\pagenumbering{roman}
\thispagestyle{empty}

\title{{\huge An Implementation Approach to Rotation Programs}

\author{Ville Korpela\thanks{Turku School of Economics, University of
Turku{\footnotesize . }E-mail: \texttt{vipeko@utu.fi}.}
\and Michele  Lombardi\thanks{Management School, University of Liverpool, Liverpool, UK. Department of Economics and Statistics, University of Napoli Federico II.} 
\and Riccardo D.  Saulle\thanks{Department of Economics and Management, University of Padova. E-mail:
\texttt{riccardo.saulle@unipd.it}}
}}

\thispagestyle{empty}
\maketitle

\thispagestyle{empty}

\begin{abstract}

\noindent We study rotation programs within the standard implementation framework under complete information. A rotation program is a myopic stable set whose states are arranged circularly, and agents can effectively move only between two consecutive states. We provide characterizing conditions for the implementation of efficient rules in rotation programs. Moreover, we show that the conditions fully characterize the class of implementable  multi-valued and efficient rules. 
 \\*[\baselineskip]

\end{abstract}

\newpage

\titleformat{\section}[hang]
{\normalfont\Large\fillast\bf\color{darkscarlet}}
{\scshape  \oldstylenums{\thesection}}
{1ex minus .1ex}{\Large}
\titlespacing{\section}{3pc}{*3}{*2}[3pc]

\titleformat{\subsection}[hang]
{\normalfont\large\fillast\bf\color{darkscarlet}}
{\scshape  \oldstylenums{\thesubsection}}
{1ex minus .1ex}{\large}
\titlespacing{\subsection}{3pc}{*3}{*2}[3pc]

\titleformat{\subsubsection}[hang]
{\normalfont\large\fillast\bf\color{darkscarlet}}
{\scshape  \oldstylenums{\thesubsubsection}}
{1ex minus .1ex}{\normalsize}
\titlespacing{\subsubsection}{3pc}{*3}{*2}[3pc]

{\hypersetup{linkcolor=maastricht}
  \tableofcontents}

\thispagestyle{empty}
\clearpage
\pagenumbering{arabic} 
\setcounter{page}{1}

\section{Introduction}
An economic department must choose a department head among its professors. However, professors would like to avoid this role due to its administrative workload. This impasse is often resolved by implementing a rotating program: each professor will take on the new task for some time. 

Rotation programs are widely used.  A prominent example is given by the business practice of job rotations, which consists of periodically rotating the jobs assigned to the employees throughout their employment. This practice has been used in many industries for a wide array of employees, from factory line workers to executives (\citet{Osterman1994, Osterman2000}, \citet{Gittleman1998}) and for different reasons.\footnote{From one side, employees who rotate accumulate more human capital because they are exposed to a broader range of
experiences. On another side, the employer itself learns more about its employees if it can observe how they perform at different jobs \citep{Arya2004}.} Furthermore, rotation programs have been practiced in managing common-pool resources as an alternative to quota and lotteries. In many areas of the world, rotating groups are formed for farming, grazing, gaining access to water, and allocating fishing spots (\citet{Ostrom1990}, \citet{Berkes1992}, \citet{Sneath1998}). Recently, \citet{Galeotti2021} show that rotation schemes can be used to prevent the spread of infections. In this view, a rotation scheme is a mechanism to shape social interactions to minimize the risk of contagion. 
Further, as illustrated by the “problem of the department head,” rotation programs can help achieve fairness in assignment problems. Indeed, we human beings tend to solve these kinds of conflicts either by using lotteries or implementing rotation schemes.
However, the literature on assignment problems focuses mainly on randomization \citep{Hofstee1990, Moulin2001, Budish2013}, though experimental evidence \citep{Eliaz2014, Andreoni2020} shows that lotteries do not avoid ex-post envy. 

In this paper, we propose an implementation approach to the study of rotation programs in which agents can rotate continuously among Pareto efficient allocations. Therefore, our challenge lies in designing a mechanism (i.e., game form) in which the behavior of agents always coincides with the recommendation given by a social choice rule (SCR). If such a mechanism exists, the SCR is implementable.

The first difficulty in adopting this approach concerns the choice of the solution concept. Most of the game-theoretical solutions used in literature, such as the core, the  (strong) Nash equilibrium, and the stable set \citep{vNM1944}, satisfy the property of  {\it internal stability}. Roughly speaking, a set of outcomes is internally stable if it is free of inner contradictions, i.e., for every outcome in the set, no agent or group can directly move to another outcome of the set and be better off. However, this property is incompatible with our objective to study how to allow rotations of desirable positions among agents. Thus, a theory of implementation in rotation programs cannot rely on solutions that satisfy internal stability.  Internal stability is relaxed in solution concepts considered modifications, extensions, or generalizations of the stable set. 
One of the most prominent is the “absorbing set.” As \citet{Inarra(2005)} point out, the notion of absorbing sets appears in the literature under different names and settings. \citet{Kalai1976} study the “admissible set” in various bargaining situations, and 
\citet{Shenoy1979} defines the “elementary dynamic solution” for coalitional games. More recently,
\citet{Inarra2013} study the absorbing set for roommate problems,  and  \citet{JacksonWatts2002} study the “closed cycle” for network formation.  
Finally, the myopic stable set (MSS), defined by \citet{DemuynckHeringsSaulleSeel2019a} for a general class of games,  includes all previous notions of absorbing sets. The MSS is the smaller set of states such that the following properties are satisfied. 1) There are no profitable deviations from a state   {\bf inside} the set to a state {\bf outside} the set. 2) For each state outside the set, a sequence of agents’ deviations converge to the set. Thus,  the MSS is a valid prediction of agents' play, though it violates internal stability because it allows deviations within the set. Furthermore, the prediction offered by the MSS is robust in the following terms:  Though agents may reach an agreement on a state outside the set, a sequence of myopic improvements will bring them back to the MSS. For this reason, we adopt the MSS as our solution concept. 

From a methodological point of view, we exploit a novel implementation technique, named implementation via {\it rights structures} (\Cref{sec2}),  recently introduced by   \citet{KorayYildiz2018}. A rights structure formalizes power distribution within society. Thus, differently from canonical mechanism design, our design exercise consists of allocating rights to agents such that their behavior always coincides with the recommendation given by an SCR. We follow this approach for two reasons. Firstly, rights structures are a generalization of effectivity functions, which are at the heart of the definition of MSSs.  From an implementation viewpoint, the effectivity relationship is the design variable, playing the role of the mechanism. Secondly, though rights structures do not model time, they effectively describe all the paths generated by agents' interactions.

We show that {\it indirect monotonicity} is sufficient for implementation in MSS via a finite rights structure.\footnote{A finite rights structure is a rights structure in which the set of states is finite.} {\it Indirect monotonicity} is weaker than Maskin monotonicity. Since this result is obtained by constructing a finite rights structure, it encompasses implementation in core and generalized stable sets \citep{vanDeemen1991, Wooders2009}. Moreover, for marriage problems  \citep{Knuth1976} and a class of exchange economies with property rights \citep{BalbuzanovKotowski2019}, we show that the set of stable outcomes is implementable in MSS. It is worth stressing here that this implementation is obtained by devising a rights structure endowed with well-defined convergence properties. Convergence is an aspect that is particularly important in our design framework. 

However,  implementation in MSS cannot always guarantee the order of rotation. Indeed, it cannot exclude the possibility that a rotation gets stuck in a cycle which rules out some agents from the process. To solve this drawback,  \Cref{rotation program} introduces the notion of \textit{implementation in rotation programs}. Implementation in rotation programs is a particular kind of implementation in MSS, in which every cycle generated within the MSS needs to be a rotation scheme. 

We identify a necessary condition, named {\it rotation monotonicity},  for implementation in rotation programs of efficient SCRs. When a multi-valued SCR describes the planner's goal, {\it rotation monotonicity} fully characterizes the class of implementable SCRs.\footnote{See, for instance,  \citet{Arunava2019}.}  Finally, \Cref{sectJobRotat} study two classes of assignment problems that implementable in rotation programs. Assignment problems in which agents share the same best/worst outcome, and assignment problems in which the planner knows that two agents have the same top-outcome.  All proofs are relegated in the \hyperlink{appendix}{Appendix B}.

\subsection*{Related Literature}
To the best of our knowledge, we are the first to study the economic design of rotation programs in an implementation framework that allows agents to rotate among Pareto efficient allocations continuously. Previous contributions consider a different notion of rotation scheme, which reduces to a one-period exchange of agents’ tasks. Indeed,  \citet{Yu2020}  and  \citet{Yu2020b} study classes of job rotations according to which employees do not necessarily circulate through tasks. Moreover, in contrast to \citet{Yu2020}  and  \citet{Yu2020b}, we focus on Pareto efficient allocations. 
  
Our contribution is also in line with \citet{Arya2004}, who study job rotations within a principal-agent framework. In particular, they identify conditions under which job rotation and specialization are each optimal. In contrast to us, their job rotation scheme does not guarantee the circulation of employees through jobs. 

Finally, our paper contributes to the literature on implementation via rights structure \citep{KorayYildiz2018,KorayYildiz2019,Korpela2019,Korpela2020} and it is broadly related to the literature on assignments problems \citep{ShapleyShubik1971, Roth1990, Abdulkadiroglu1998}.

\section{The Setup}
\label{sec2}
We consider a finite (nonempty) set of \textit{agents}, denoted by
$N$, and a finite (nonempty) set of \textit{alternatives},
denoted by $Z$. We endow $Z$ with a metric $\hat{d}$.  For every set $A$, the power set of $A$ is denoted by
$\mathcal{A}$ and $\mathcal{A}_{0}\equiv\mathcal{A-\{\varnothing\}}$ is the
set of all nonempty subsets of $A$. Each element $K$ of $\mathcal{N}_{0}$ is
called a \textit{coalition}.  A \textit{preference ordering} $R_{i}$ is a
complete and transitive binary relation over $Z$. Each agent $i$($\in N$) has
a preference ordering $R_{i}$ over $Z$. The \textit{asymmetric} part $P_{i}$ of $R_{i}$ is defined by $xP_{i}y$ if and only if $xR_{i}y$ and not $yR_{i}x$, while the \textit{symmetric} part $I_{i}$ of $R_{i}$ is defined by $xI_{i}y$
if and only if $xR_{i}y$ and $yR_{i}x$. A \textit{preference profile} is thus
an $n$-tuple of preference orderings $R\equiv\left(  R_{i}\right)  _{i\in N}$.
For any profile $R$ and $K \in \mathcal{N}_{0}$, we write $xR_{K}y$ to denote that $xR_{i}y$ holds for all $i\in K$ and
$xP_{K}y$ to denote that $xP_{i}y$ holds for all $i\in K$.
As usual, $L_i(x,R)$ denotes the lower contour set of $x$ at $R$ for agent $i$. 
The \textit{preference domain}, denoted by $\mathcal{R}$, consists of the set of admissible preference profiles satisfying the following property: 
\begin{equation}
 R\in \mathcal{R\iff }\text{for all }x,y\in Z:\text{ if }xI_{N}y\text{, then }x=y\text{. } \label{pref-dom}
\end{equation}

The domain of preferences underlying classical assignment problems, which are our main focus, satisfies the above property. 

The goal of the planner is to implement a \textit{social choice rule} (SCR) $F$, defined by  $F:\mathcal{R}%
\longrightarrow\mathcal{Z}_{0}$. We refer to $x\in F\left(  R\right)  $ as an
$F$-optimal outcome at $R$. The \textit{range} of $F$ is the set
\[
F\left(  \mathcal{R}\right)  \equiv\left\{  x\in Z|x\in F\left(  R\right)
\text{ for some }R\in\mathcal{R}\right\}  \text{.}%
\]

\noindent The \textit{graph} of $F$ is the set 
\[
Gr(F)\equiv\{(x,R)|x\in F(R), R\in {\cal R}\}
\]

\noindent We impose  the following  assumption on $F$:

\begin{definition}[{\it Efficiency}]
We say that SCR $F$ is \emph{efficient}, if for all $R \in \mathcal{R}$, and all $z \in F(R)$, there does not exist any $x \in Z$ such that $x R_{N} z$ and $x P_{i} z$ for at least one agent $i \in N$.
\end{definition}

To present our theory we find convenient to move away from  canonical mechanism. Thus, we  rely on a particular kind of implementation framework which models rights  distribution  within the society.
Roughly speaking, we assume that  a planner first describes the available alternatives via a set of possible states. Then, he specifies which agent or group has the right to move from a state to another. The rights distribution is    such that, for any state of the world, the prediction of the solution concept returns the socially desirable alternatives. Formally,  
to implement $F$, the  
 planner constructs a \textit{rights
structure} $\mathnormal{\Gamma}=\left((S,d),h,\gamma\right)  $, where
$S$ is the \textit{state space} equipped with a metric $d$, $h:S\rightarrow Z$ the \textit{outcome
function}, and $\gamma$ a \textit{code of rights}, which is a (possibly empty)
correspondence $\gamma:S\times S\twoheadrightarrow\mathcal{N}$. Subsequently,
a code of rights specifies, for each pair of distinct states $\left(
s,t\right)  $, the family of coalitions $\gamma\left(  s,t\right) \subseteq \mathcal{N}$ that is  entitled to
move  from state $s$ to $t$. If $\gamma(s,t)=\emptyset$ then no coalition is entitled to move from $s$ to $t$. The rights structure $\mathnormal{\Gamma}$ presented here  is an augmented version of  the  right structure previously introduced by \citet{KorayYildiz2018} which does not  includes  the metric $d$. 
From an 
economic design perspective, the  right structure  is the planner's design variable and
corresponds to a ``mechanism'' in the economic theory jargon.   A rights structure $\Gamma$ is said to be an individual-based rights structure if, for each pair of distinct states $(s,t), \gamma(s,t)$ contains only unit coalitions if it is nonempty. A rights structure $\Gamma$ is termed finite if the state space $S$ is a finite set.

A right structure together with a preference profile returns a  \textit{social environment} \citep{Chwe1994}, a general  framework  to model strategic interaction among agents or groups.

\begin{definition}[Social Environment]
A \textit{social environment}  is    a pair $(\Gamma, R)$ consisting of a  right structure $\Gamma$ together with a preference profile $R$.
\end{definition}

Next, a model of behavior is needed to predict at what state the agents are going to end up with. 
This is often done by selecting an equilibrium concept. A common and unifying way
that resonates across all microeconomics is to use the \emph{core} defined in terms of strong domination.

\begin{definition}[{\it Core}]
\label{def2}
 For any social environment $(\Gamma, R)$, 
a state $s\in S$ is an \emph{core element} at $R$ if $h\left(  t\right)  P_{K}h\left(
s\right)$ does not hold for any $t\in S$ and 
$K\in\gamma\left(  s,t\right)  $. We write $C\left(  \mathnormal{\Gamma},R\right)  $ for the set of core elements at $R$.
\end{definition}

\citet{KorayYildiz2018} study implementation problem in core\footnote{The notion of $\Gamma$-equilibrium provided by \citet{KorayYildiz2018} is equivalent to the notion of core for social environment \citet{DemuynckHeringsSaulleSeel2019a} employed here.} via rights structures.\footnote{\citet{Korpela2020} provide a full characterization of the class of implementable SCRs.} To speak,  an SCR is implementable in core  by a finite right structure if, at any preference profile,  the outcomes induced by any core element are those  demeed socially optimal and {\it vice versa}.

\begin{definition}[{\it Implementation in core}]

  A rights structure $\mathnormal{\Gamma}$ implements $F$ in core
if $F\left(  R\right)  =h\circ C\left(  \mathnormal{\Gamma},R\right)  $ holds for all
$R\in\mathcal{R}$. If such a rights structure exists, $F$ is
implementable in core by a rights structure.
\end{definition}

\section {Towards   Implementation In  Rotation\\ Programs} \label{sect untermediate step}
As outlined above, the fundamental idea of our notion of  implementation in rotation programs relies on the   Myopic Stable Set (MSS) \citep{DemuynckHeringsSaulleSeel2019a}. 
As a first step, this section presents the MSS and  studies its  implementation  via rights structures.

\subsection{Implementation In  Myopic Stable Set} 

To define the MSS, we need the notion of a {\it myopic improvement path}.\footnote{If the state space is finite then \Cref{def4} reduces to the following:
A sequence of states $s_{1}, \ldots,s_{m}$ is called a \emph{myopic improvement path} from state $s_{1}$ to set $T \subseteq S$ at $R$, if $s_{m} \in T$, and there exists a collection of coalitions $K_{1}, \ldots,K_{m-1}$ such that, for  $j=1, \ldots,m-1,$, (i) $K_{j} \in \gamma(s_{j},s_{j+1})$ and (ii) $h(s_{j+1})P_{K_{j}}h(s_{j})$. 
} There is a myopic improvement path from a state $s$ to a set $T$ if a sequence of coalitional deviations from $s$ to a state arbitrarily close to $T$ exists such that every coalition involved in the sequence has the power as well as the incentive to move.

\begin{definition}[{\it Myopic Improvement Path}]
\label{def4}
Given a social environment
$(\mathnormal{\Gamma}, R)$,  
   a sequence of states $s_{1}, \ldots,s_{m}$ is called a myopic improvement path from state $s_{1}$ to set $T \subseteq S$ at $R$, if  for all $\epsilon>0$ there exists a state $s \in T$ such that $d(s,s_m)<\epsilon$  and   a collection of coalitions $K_{1}, \ldots,K_{m-1}$ such that, for $j=1, \ldots,m-1,$
\smallskip

(i) $K_{j} \in \gamma(s_{j},s_{j+1})$  
\smallskip

(ii) $h(s_{j+1})P_{K_{j}}h(s_{j})$
\end{definition}

The MSS can be defined as follows:\footnote{ \noindent When the set of states is finite, Condition 2 reduces to the following one:
Iterated External stability:  For all $t \in S \setminus M$, there exists a finite myopic improvement path from $t$ to $M$.
}

\begin {definition}[{\it Myopic Stable Set}] 
\label{def5} 
The set $mss(\Gamma, R) \subseteq S$ is an MSS  at 
$(\mathnormal{\Gamma}, R)$ if it is {\cal closed} and  satisfies the following three conditions:
\begin{enumerate}

\item \emph{Deterrence of external deviations}: For all $s \in mss(\Gamma, R)$, and all $t \in S \setminus mss(\Gamma, R)$, there is no coalition $K \in \gamma(s,t)$, such that $h\left(  t\right)  P_{K}h\left(s\right)$.
\item \emph{Asymptotic external stability}: For all $t \in S \setminus mss(\Gamma, R)$, there exists a myopic improvement path from $t$ to $mss(\Gamma, R)$.
\item \emph{Minimality}: There is no set $M' \subset mss(\Gamma, R)$ that satisfies the two conditions above.
\end{enumerate}  
\end{definition}
 
\emph{Deterrence of external deviations} requires that from any state in the set, there are no coalitional deviations to states outside the set. \emph{Asymptotic external stability} states a myopic improvement path to the set exists from any state outside the set. Finally, \emph{Minimality} requires that the MSS is the smaller closed set satisfying the first two conditions.

Let MSS($\mathnormal{\Gamma}$, $R$)=$\{s \in S \mid s\in mss(\Gamma, R)\}$ be the union of all MSSs at $(\Gamma, R)$.
We are now ready to introduce our notion of implementation in MSS: an SCR is implementable in MSS by a finite rights structure if, for each preference profile, the outcomes selected by $F$ coincide with those of the MSS.

\begin{definition}[{\it Implementation in MSS}]
\label{def6}
 A rights structure $\mathnormal{\Gamma}$ implements $F$ in MSS
if $F\left(  R\right)  =h\circ MSS\left(  \mathnormal{\Gamma},R\right)  $ holds for all
$R\in\mathcal{R}$. If such a rights structure exists, $F$ is
implementable in MSS by a rights structure. 
\end{definition}

We will be using the following sufficient condition in our characterization result. 

\begin{definition}[{\it Indirect Monotonicity}]

An SCR $F$ satisfies {\it indirect monotonicity} provided that  for all $R,R'\in \mathcal{R}$, and all $z \in Z$, 
if $z\in F(R)$ and $z\notin F(R')$ with $L_i(z,R)\subseteq L_i(z,R')$ for all $i \in N$, then there exist a  sequence of outcomes  $\{z_{1}, \ldots,z_{h}\}\subseteq F(R)$  with $z=z_{1}$, $z\neq z_h$ and  a sequence of agents $i_{1},\ldots,i_{h-1}$ such that:
\smallskip

(i) $z_{k+1}  P'_{i_{k}} z_{k}$ \ for all $k \in \{1, \ldots,h-1\}$
\smallskip

(ii) $L_i(z_h,R)\not\subseteq L_i(z_h,R')$ for some $i \in N$.

\end{definition}

Suppose that $z$ is $F$-optimal at $R$. Further, suppose that preferences change from $R$  to $R^{\prime }$ in such a way the standing of $z$ improves for every agent. Finally, suppose that $z$ is not $F$-optimal at $R^{\prime }$. Then,
{\it indirect monotonicity} says that there exist a agent $i$ and a pair of
outcomes $(z^{\ast },y)$ such that $y$ improves with respect to $z^{\ast }$
for agent $i$ when preferences change from $R$ to $R^{\prime }$ (i.e., there
is a preference reversal), where $z^{\ast }$ is $F$-optimal at $R$ and $z$
is connected to $z^{\ast }$ via a "myopic improvement path" at $R^{\prime }
$ involving only $F$-optimal outcomes at $R$. 

The latter requirement
differentiates {\it indirect monotonicity} from Condition $\alpha $ of \cite{AbreuSen90}, according
to which no outcome of the sequence has to be $F$-optimal.
{\it Indirect monotonicity} is implied by (Maskin) monotonicity,  and they are equivalent when $F$ is single-valued. Monotonicity says that if an outcome $z$ is $F$-optimal at the profile $R$ and this $z$ does not strictly fall in preference for anyone when the profile changes to $R^{\prime}$, then $z$ must remain a $F$-optimal outcome at $R^{\prime}$.
The following result characterizes a class of implementable SCRs in MSS by a finite rights structure.\footnote{When $Z$ is not a finite set, by using the rights structure designed in the proof of \Cref{Th1}, it is possible to show that it implements $F$ in MSS when $F$ is closed valued and upper hemi-continuous,  the set of alternatives $Z$ is compact and the domain $\mathcal{R}$ is also compact.}

\begin{theorem}
\label{Th1}
Any  efficient $F$ satisfying {\it indirect monotonicity} is implementable in MSS by a finite rights structure.
\end{theorem}

{\it Indirect monotonicity} is a sufficient condition for implementation in MSS via rights structures,  though it is not necessary. Example 1 in \citet{Korpela2021} makes the point.

Before presenting implementation in rotation programs, we discuss in the following subsections the importance of  \Cref{Th1}. However, the impatient reader can move to \Cref{rotation program}  without loss of understanding.

\subsection{Convergence Property}

As \citet{Jackson1992} and \citet{Moore1992} point out, canonical mechanisms for implementing socially desirable outcomes have unnatural futures: they are highly complex and challenging to explain in natural terms. In particular, when agents are boundedly rational, such mechanisms may lead to the convergence of undesirable outcomes.
Our result shows that even unsophisticated agents, using elementary adjustment rules, can reach desirable outcomes; our mechanism is robust to some bounded rationality.
Indeed, \Cref{Th1} demonstrates that the
implementing rights structure guarantees the convergence to a myopic stable
state in a finite number of transitions among states. The reason is that
our implementation problems are solved by devising a finite rights structure. This property assures that the MSS can be reached in a finite sequence of myopic improvements from any state outside it.

\begin{corollary}
\label{cor1}Every efficient and monotonic $F:\mathcal{R}%
\longrightarrow\mathcal{Z}_{0}$ is
implementable in MSS via a finite rights structure.
\end{corollary}

This result can be thought of as the counterpart of recurrent implementation in better-response dynamics studied by \citet{CabralesSerrano2011}, in which agents myopically adjust their actions in the direction of better-responses. When combined with a “no-worst-alternative condition,” these authors show that a variant of monotonicity is a key condition for implementation in recurrent strategies.  \Cref{cor1} shows that for assignment problems of indivisible goods, monotonicity, together with Pareto efficiency, is sufficient for a similar type of implementability.

In \hyperlink{appendixA}{Appendix A}, we study two models where convergence is desirable. In particular, we consider exchange economies with complex endowment systems recently introduced by \citet{BalbuzanovKotowski2019} as well as the class “pure marriage problems” studied by \citet{Knuth1976}. Both models do not satisfy any converge property. We show that the direct exclusion core of  \citet{BalbuzanovKotowski2019} and the solution that selects all stable matchings in the sense of \citet{Knuth1976}, can be implemented in MSS.

\subsection{Connections To Other Implementability Notions}

We conclude this section by showing that implementation in MSS by a finite rights structure is equivalent to implementation in absorbing set and implementation in generalized stable set \citep{vanDeemen1991, Wooders2009}. Before showing it, let us formally introduce these alternative notions of stability.

\begin{definition}[Absorbing Set] 
\label{def10} Let us assume that $S$ is finite. The set $A(\Gamma, R)\subseteq S$ is an absorbing set at $(\Gamma, R)$ if it satisfies the following two conditions:

\noindent {\bf (\textit{a})} For all $s,t\in A(\Gamma ,R)$, there exists a finite
myopic improvement path from $t$ to $s$.

\noindent {\bf(\textit{b})} For all $t\in S\backslash A(\Gamma ,R)$ and $s\in A(\Gamma ,R$), there does not exist any finite myopic improvement path from $s
$ to $t$.

\end{definition}

Let ${\cal A}\left(  \mathnormal{\Gamma},R\right)$ be the union of all absorbing sets at $(\Gamma, R)$. 
The following establishes the notion of implementation in absorbing set.

\begin{definition}[{\it Implementation in Absorbing Sets}]
\label{def11}
 A rights structure $\mathnormal{\Gamma}$ implements $F$ in absorbing set
if $F\left(  R\right)  =h\circ {\cal A}\left(  \mathnormal{\Gamma},R\right)  $ for all
$R\in\mathcal{R}$. If such a rights structure exists, $F$ is
implementable in absorbing sets by a rights structure. 
\end{definition}

\begin{definition}[{\it Generalized Stable  Set}] 
\label{def12} Let us assume that $S$ is finite. The set $V(\Gamma, R)\subseteq S$ is a generalized stable set at $(\Gamma, R)$ if it satisfies the following two conditions:

\noindent {\bf 1.} {\it Iterated Internal Stability:} For all $s,t\in V(\Gamma ,R)$, there is no   finite
myopic improvement paths from $t$ to $s$.

\noindent {\bf 2.} {\it Iterated External  Stability:} For all $t\in S\backslash V(\Gamma ,R)$ there exists  a finite myopic improvement path from $t$ to $V$.

\end{definition}

Let ${\cal V}\left(  \mathnormal{\Gamma},R\right)$ be the union of all generalized stable sets at $(\Gamma, R)$. 
As usual, we establishes the notion of implementation in generalized stable set.

\begin{definition}[{\it Implementation in Generalized Stable Set}]
 A rights structure $\mathnormal{\Gamma}$ implements $F$ in generalized stable set
if $F\left(  R\right)  =h\circ {\cal V}\left(  \mathnormal{\Gamma},R\right)  $ for all
$R\in\mathcal{R}$. If such a rights structure exists, $F$ is
implementable in generalized stable set by a rights structure. 
\end{definition}

\citet{Inarra(2005)} and \citet{Nicolas2009} study the relation between absorbing sets and generalized stable sets. \citet{Korpela2021} (Theorem 2) provide further insights into the relationship between these solution concepts. In particular, they show that when the state space is finite, the union of generalized stable sets is equivalent to the union of absorbing sets, which, in turn, is equivalent to the unique myopic stable set. 

 \Cref{Th1}, when combined with  Theorem 2 in \citet{Korpela2021}, gives us the following significant result.\footnote{The proof of \Cref{cor4} is omitted.}

\begin{corollary}
\label{cor4}
Any efficient $F$ satisfying {\it indirect monotonicity} 
is implementable in absorbing sets by a finite right structure, and in generalized stable sets by a finite right structure.
\end{corollary}

\section{Rotation Programs}
\label{rotation program}

As noted earlier,  implementation in MSS is only a preliminary step towards implementation in rotation programs. Indeed, on the one hand, implementation in MSS gives the planner the ability to design cycles among socially optimal outcomes. On the other hand, the planner does not have complete control of the cycles, in the sense that it cannot always guarantee that all agents circulate through all socially optimal outcomes. We illustrate this point through the following example.

\begin{example}
\label{ex2}

Suppose that $N=\{1,2,3\}$, $Z=\{x,y,z\}$, and $\mathcal{R}=\{R,R'\}$. Preference are defined in the table below.

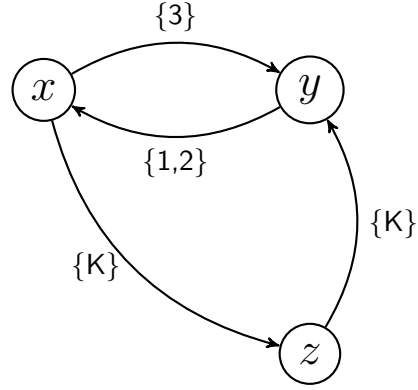
\begin{figure}[h]
\centering
\begin{minipage}{0.3\textwidth}
\begin{tabular}
[c]{c|c|c|c|c|c}%
\multicolumn{3}{c|}{$R$} & \multicolumn{3}{|c}{$R^{\prime}$}\\\hline
\emph{1} & \emph{2} & \emph{3} & \emph{1} & \emph{2} & \emph{3}\\\hline
$x$ & $z$ & $y$ & $x$ & $x$ & $y$ \\
$y$ & $x$ & $z$ & $y$ & $y$ & $x$ \\
$z$ & $y$ & $x$ & $z$ & $z$ & $z$
\end{tabular}
\end{minipage}
\hspace{1.5cm}
\begin{minipage}{0.3\textwidth}

\begin{tikzpicture}[->,>=stealth',auto,node distance=3.5cm,
  thick,main node/.style={circle,draw,font=\sffamily\Large\bfseries}]

  \node[main node] (1) {$x$};
  \node[main node] (2) [right of=1] {$y$};
  \node[main node] (3) [below of=2] {$z$};

  \path[every node/.style={font=\sffamily\small}]
    (1) edge [->, bend left] node  [above] {\{3\}} (2)
    (1) edge [<-, bend right] node  [below] {\{1,2\}} (2)
    (2) edge [<-, bend left] node [right] {\{K\}} (3)
    (3) edge [<-, bend left] node [left] {\{K\}} (1);

\end{tikzpicture}

\end{minipage}

\caption{Preferences and  implementing rights structure. $\#K\geq 2$}\label{figure_IM}
\end{figure}

Let $F$ be such that $F(R)=\{x,y,z\}$ and $F(R')=\{x,y\}$. This SCR satisfies {\it indirect monotonicity} because $F(R') \subseteq F(R)$, $F(R) \setminus F(R')=\{z\}$ and $L_{3}(z,R) \nsubseteq L_{3}(z,R')$. Note that at $R'$ only agent $3$ wants to move from $x$ to $y$, and agents $1$ and $2$ want to move from $y$ to $x$. Therefore, to produce a rotation among  $\{x,y\}$ at  $R'$, it is necessary to give to agent $3$ the power to move from $x$ to $y$ and to agent $1$ or $2$ the power to move from $y$ to $x$. A rights structure that implements $F$ in MSS is depicted in  \Cref{figure_IM}, in which the set of states is $S=Z$, the outcome function is the identity map, and in which $\gamma$ is represented by the arrows.
Note that at $R$, such a rights structure generates a sub-cycle in which the outcome $z$ is ruled out. Consequently,  the rotation among states $\{x,y,z\}$ cannot be guaranteed.  
\end{example}

We solve this drawback by focusing on a subset of MSS.

\subsection{Implementation In  Rotation Programs}

We start by defining   a rotation program as follows.

\begin{definition}[{\it Rotation Program}]
\label{def14}A {\it rotation program} for $\left( \Gamma ,R\right) $
is an ordered subset of states $\bar{S}=\left\{ s_{1},...,s_{m}\right\}
\subseteq S$ such that for all $s_{i},s_{i+1}\in \bar{S}$:

\begin{description}
\item[(\textit{i})] For all $s\in \bar{S}\backslash \left\{ s_{i}\right\} $,
$h\left( s_{i}\right) \neq h\left( s\right) $.

\item[(\textit{ii})] For all $s\in S\backslash \left\{ s_{i},s_{i+1}\right\}
$ and all $K\in \mathcal{N}_{0}$, if $K\in \gamma \left(
s_{i},s\right) $, then not $h\left( s\right) P_{K}h\left( s_{i}\right) $.

\item[(\textit{iii})] There exists $K\in \mathcal{N}_{0}$ such that $K\in
\gamma \left( s_{i},s_{i+1}\right) $ and $h\left( s_{i}\right) P_{K}h\left(
s_{i+1}\right) $.
\end{description}
\end{definition}

Condition (i) states that in a rotation program there are no two states providing the same outcome; conditions (ii) and (iii) together  require that the only possible transition occurs among adjacent states. 
Our notion of implementation in rotation programs can be stated as follows.

\begin{definition}[{\it Implementation in Rotation Programs}]
\label{def15}A rights structure $\Gamma $ implements $F$ in rotation programs if the following requirements are satisfied:

\begin{description}
\item[(\textit{i})] $\Gamma $ implements $F$ in MSS.

\item[(\textit{ii})] For all $R\in \mathcal{R}$, $MSS\left( \Gamma ,R\right) $ is partitioned in
rotation programs $\left\{ S_{1},...,S_{m}\right\} $ such that $h\circ
S_{i}=F\left( R\right) $ for all $i=1,...,m$.
\end{description}

\noindent If such a rights structure exists, we say that $F$ is  {\it implementable in rotation programs}.
\end{definition}

Roughly speaking, the above notion of implementation refines our notion of implementation in MSS, in the sense all myopic stable states must be arranged circularly. Thus, and irrespective of agents' preferences, the core of an implementing rights structure is always empty when $F(R)$ has more than one outcome.

\subsection{Characterization Results}

In what follows, we introduce the notion of {\it rotation monotonicity}, which is at the hearth of the theory we develop here.

\begin{definition}[{\it Rotation Monotonicity}]
\label{def16}

$F:\mathcal{R\longrightarrow Z}_{0}$ satisfies {\it rotation monotonicity} provided that for all $R\in \mathcal{R}$, elements of $F\left( R\right) $
can be ordered as $x\left( 1,R\right) ,...,x\left( m,R\right) $ for some
integer $m\geq 1$ such that for all $R^{\prime }\in \mathcal{R}$, if $%
F\left( R\right) \neq F\left( R^{\prime }\right) $ and either $\#F\left(
R^{\prime }\right) >1$ or $\left[ \#F\left( R^{\prime }\right) =1\text{ and }%
F\left( R^{\prime }\right) \notin F\left( R\right) \right] $, then for each $x\left(
i,R\right) \in F\left( R\right) $, there exist a sequence of agents $%
i_{1},...,i_{h}$, states $\{x(i,R),x(i+1,R),...,x(i+h,R)\}\subseteq F(R)$ with $1\leq h\leq m$ and an outcome $z\in Z$  such that
\medskip

    \noindent $\bullet \ 
x\left( i+\ell +1,R\right) P_{i_{\ell +1}}^{\prime }x\left( i+\ell ,R\right)\ \ \ 
\forall \ell \in \left\{ 0,...,h-1\right\} 
$ 
\medskip

\noindent  $\bullet \ 
x\left( i+h,R\right) R_{i_{h}}z \ \ \text{ and }\ \ zP_{i_{h}}^{\prime }x\left(
i+h,R\right)$.

\end{definition}

When preferences change from $R$ to $R^{\prime }$ and $%
F\left( R\right) \neq F\left( R^{\prime }\right) $, {\it rotation monotonicity} requires that for
every $z$ which  is $F$-optimal at $R$, there is an agent $i$ and a pair $%
\left( z^{\ast },y\right) $ such that: (i) $y$ improves with respect to $z^{\ast }
$ for agent $i$ as preferences change; (ii)   $z^{\ast }$  is $F$-optimal at $%
R$ and  it s connected to $z$ via a specific  “myopic improvement path" at $%
R^{\prime }$ that not only involves just $F$-optimal outcomes at $R$ but
also obeys the circular arrangement of the elements of $F\left( R\right) $.

The above property implies {\it indirect monotonicity} when $\#F\left( R\right)
\neq 1$ for all $R\in \mathcal{R}$. With respect to {\it indirect monotonicity},
{\it rotation monotonicity} requires that all $F$-optimal outcomes at $R$ must be arranged
circularly. 
The next result shows that only SCRs satisfying {\it rotation monotonicity} are implementable in rotation programs.

\begin{theorem}[{\it Necessity}]
\label{Th3}If $F$ is implementable in rotation programs, then it satisfies {\it rotation monotonicity}.
\end{theorem}

Recall that the SCR in   \Cref{ex2}  is not implementable in rotation programs.
It is illustrative to  study the SCR of the example   in the light of \Cref{Th3}.
\begin{example}[continues=ex2]
The social choice rule $F$ in \Cref{ex2} does not satisfy {\it rotation monotonicity}. To see this, notice that there are two cyclic orderings of $F(R)$ $-$ $x,y,z$ and $x,z,y$. Both violate {\it rotation monotonicity}. Ordering $x,y,z$ violates {\it rotation monotonicity} because $L_{i}(y,R) \subseteq L_{i}(y,R')$ and $z \in L_{i}(y,R')$ for all $i \in N$, and $x,z,y$ violates {\it rotation monotonicity} because $L_{i}(x,R) \subseteq L_{i}(x,R')$ and $z \in L_{i}(x,R')$ for all $i \in N$.    
\end{example}

Observe,  that {\it rotation monotonicity} has a bite only  when either $\#F\left( R^{\prime }\right)
>1$ or [$\#F\left( R^{\prime }\right) =1$ but $F\left( R^{\prime }\right)
\notin F\left( R\right) $] and it is vacuously satisfied otherwise.
It follows that  {\it rotation monotonicity} alone is not a sufficient condition for implementation in rotation programs. However, we show that   it is sufficient together with another auxiliary condition termed {\it Property M}, which can be defined as follows.

\begin{definition}[{\it Property M}]

$F:\mathcal{R\longrightarrow Z}_{0}$ satisfies {\it Property M} provided that for all $R\in \mathcal{R}$, elements of $F\left( R\right) $
can be ordered as $x\left( 1,R\right) ,...,x\left( m,R\right) $ for some
integer $m\geq 1$ such that for all $R^{\prime }\in \mathcal{R}$, if $%
F\left( R\right) \neq F\left( R^{\prime }\right) $ and $F\left( R^{\prime }\right) =\left\{ x\left(
k,R\right) \right\} $ for some $1\leq k\leq m$, then either 
\medskip

    \noindent $\bullet$   the conclusion
of  {\it rotation monotonicity} holds for all $x\left( j,R\right) \in F\left( R\right) \backslash \left\{ x\left( k,R\right) \right\} $ 
\medskip

    \noindent $\bullet$ or for  
each $x\left( j,R\right) \in F\left( R\right) \backslash \left\{ x\left( k,R\right) \right\}  $ for which the conclusion of {\it rotation monotonicity} does not hold,
there exists a sequence of agents $i_{1},...,i_{\ell }$ such that%
\[
x\left( k,R\right) P_{i_{\ell }}^{\prime }x\left( k-1,R\right) P_{i_{\ell
-1}}^{\prime }\cdot \cdot \cdot P_{i_{2}}^{\prime }x\left( j+1,R\right)
P_{i_{1}}^{\prime }x\left( j,R\right) \ \text{and}
\]
\[L_{i}\left( x\left( k,R\right) ,R\right) \cup \left\{ x\left(
k+1,R\right) \right\} \subseteq L_{i}\left( x\left( k,R\right) ,R^{\prime
}\right)  \ \ \forall i\in N.\]

\end{definition}

\begin{theorem}[{\it Sufficiency}]
\label{Th4}If $F$ is efficient and it satisfies
{\it rotation monotonicity} and {\it Property M} with respect to the same ordered set of outcomes in $F(R)$, for all $R \in \mathcal{R}$, then it is implementable in rotation programs by a finite rights structure.
\end{theorem}

We conclude this section by considering the case that a multi-valued SCR describes the planner's goal at any preference profile. As discussed by \citet{Arunava2019}, this is a relevant case. Under these circumstances, since {\it Property M} is always satisfied, {\it rotation monotonicity} fully characterizes the class of implementable rules in rotation programs. The following result establishes the point.\footnote{The proof of \Cref{cor5} is omitted.}

\begin{corollary} 
\label{cor5}
Suppose $\#F(R)>1$ for all $R\in \cal{R}$. Then $F$ is implementable in rotation programs if and only if $F$ satisfies {\it rotation monotonicity}.  

\end{corollary}

\section{Assignment Problems}\label{sectJobRotat}

A basic yet widely applicable problem in economics is to allocate indivisible objects to agents. This problem is referred to as
the assignment problem. In this setting, there is a set of objects, which we term as “jobs", and the goal is to allocate them among the agents in an optimal manner without allowing transfers of money. The assignment problem is a fundamental setting that is not an economic environment. Since the model applies to many resource allocation settings in which the objects can be public houses, school seats, course enrollments, car park spaces, chores, joint assets of a divorcing couple, or time slots in
schedules, we now apply \Cref{cor5} to this fundamental setting.

A job rotation problem $\left( N,J,P\right) $ is a triplet where $N=\left\{ 1,...,n\right\} $ is a finite set of agents with $n\geq 2$,  $J=\left\{ j_{1},...,j_{n}\right\} $ is a finite set of jobs, $P=(P_{i})_{i\in N}$ is a profile of linear orderings such that every $%
P_{i}\subseteq J\times J$. Let $\left( N,J,P\right) $ be a job rotation problem. Every agent $i$'s
preferences over $J$ at $P_{i}$ can be extended to an ordering over the set of allocations 
$\bar{J}=\left\{ j\in J^{n}|j_{k}\neq j_{l}\text{ for all }k,l\in N\right\}$ 
in the following natural way:%
\[
jR_{i}j^{\prime }\Leftrightarrow \text{either }j_{i}P_{i}j_{i}^{\prime }%
\text{ or }j_{i}=j_{i}^{\prime }\text{,\hspace*{4mm}for all }j,j^{\prime
}\in \bar{J}\text{.}
\]%
Let $\mathcal{R}$
denote the set of all (extended) preference profiles.
The following example shows that no every efficient $F$ on $\mathcal{R}$ is
implementable in rotation programs.

\begin{example}
\label{ex4}
Let $F$ be the efficient SCR defined over $\mathcal{R}$. Suppose that there
are three agents. Let the profiles $P,P^{\prime },P^{\prime \prime } $ be defined
as follows:%
\begin{center}

\begin{tabular}{lll}
& $P$ &  \\ \hline
1 & \multicolumn{1}{|l}{2} & \multicolumn{1}{|l}{3} \\ \hline
$j_{1}$ & \multicolumn{1}{|l}{$j_{1}$} & \multicolumn{1}{|l}{$j_{2}$} \\
$j_{3}$ & \multicolumn{1}{|l}{$j_{2}$} & \multicolumn{1}{|l}{$j_{3}$} \\
$j_{2}$ & \multicolumn{1}{|l}{$j_{3}$} & \multicolumn{1}{|l}{$j_{1}$}%
\end{tabular}%
\text{ , }%
\begin{tabular}{lll}
& $P^{\prime }$ &  \\ \hline
1 & \multicolumn{1}{|l}{2} & \multicolumn{1}{|l}{3} \\ \hline
$j_{1}$ & \multicolumn{1}{|l}{$j_{1}$} & \multicolumn{1}{|l}{$j_{3}$} \\
$j_{3}$ & \multicolumn{1}{|l}{$j_{2}$} & \multicolumn{1}{|l}{$j_{2}$} \\
$j_{2}$ & \multicolumn{1}{|l}{$j_{3}$} & \multicolumn{1}{|l}{$j_{1}$}%
\end{tabular}%
\text{ and }%
\begin{tabular}{lll}
& $P^{\prime \prime }$ &  \\ \hline
1 & \multicolumn{1}{|l}{2} & \multicolumn{1}{|l}{3} \\ \hline
$j_{1}$ & \multicolumn{1}{|l}{$j_{1}$} & \multicolumn{1}{|l}{$j_{2}$} \\
$j_{3}$ & \multicolumn{1}{|l}{$j_{3}$} & \multicolumn{1}{|l}{$j_{3}$} \\
$j_{2}$ & \multicolumn{1}{|l}{$j_{2}$} & \multicolumn{1}{|l}{$j_{1}$}%
\end{tabular}%
\text{.}
\end{center}
\bigskip
It can easily be checked that $F\left( R\right)  =\left\{ \left( j_{3},j_{1},j_{2}\right) ,\left(
j_{1},j_{2},j_{3}\right) ,\left( j_{1},j_{3},j_{2}\right) \right\}$, $F\left( R^{\prime }\right)  =\left\{ \left( j_{3},j_{1},j_{2}\right)
,\left( j_{1},j_{2},j_{3}\right) \right\}$ and $F\left( R^{\prime \prime }\right)  =\left\{ \left(
j_{3},j_{1},j_{2}\right) ,\left( j_{1},j_{3},j_{2}\right) \right\}$. $F$ is not implementable in rotation programs because it violates
{\it rotation monotonicity}. To see it, assume, to the contrary, that $F$ satisfies {\it rotation monotonicity}. Then, the elements of $F\left( R\right) $ can be ordered as $
x\left( 1,R\right) ,x\left( 2,R\right) ,x\left( 3,R\right)  $.

Let us consider $R^{\prime \prime }$. Select $i \in N$ such that $x\left( i,R\right)
=\left( j_{3},j_{1},j_{2}\right) $. We show that $x\left(
i+1,R\right) =\left( j_{1},j_{3},j_{2}\right) $. Since $x\left( i,R\right) $
has not fallen strictly in anyone's preference ordering because $R^{\prime
\prime }$ is a monotonic transformation of $R$ at $\left(
j_{3},j_{1},j_{2}\right) =x\left( i,R\right) $--$L_{i}\left( \left(
j_{3},j_{1},j_{2}\right) ,R\right) \subseteq L_{i}\left( \left(
j_{3},j_{1},j_{2}\right) ,R^{\prime }\right) $ for each agent $i$, it
follows that we can only move to the next element of the ordered set, that
is, to $x\left( i+1,R\right) $. Since the top-ranked job for agent $2$ at $%
P^{\prime \prime }$ is $j_{1}$ and since, moreover, the top-ranked job for
agent $3$ at $P^{\prime \prime }$ is $j_{2}$, it follows that only agent 1
can move to $x\left( i+1,R\right) $ at $R^{\prime \prime }$, which implies
that $x\left( i+1,R\right) $ must coincide with $\left(
j_{1},j_{2},j_{3}\right) $, that is, we have that $x\left( i+1,R\right)
P_{1}^{\prime \prime }x\left( i,R\right) $ and $x\left( i+1,R\right) =\left(
j_{1},j_{3},j_{2}\right) $.\footnote{%
It cannot be that $x\left( i+1,R\right) =\left( j_{1},j_{3},j_{2}\right) $
because this would lead to the contradiction that $x\left( i+2,R\right)
=\left( j_{3},j_{1},j_{2}\right) $. The reason is that there cannot be any
preference reversal around $\left( j_{1},j_{2},j_{3}\right) $ because $%
R^{\prime \prime }$ is a monotonic transformation of $R$ at $\left(
j_{1},j_{3},j_{2}\right) $. Thus, we can only move to next element of the
ordered set. Since the top-ranked job for agent $1$ at $P^{\prime \prime }$
is $j_{1}$ and since, moreover, the top-ranked job for agent $3$ at $%
P^{\prime \prime }$ is $j_{2}$, the allocation $x\left( i+2,R\right) $ must
coincide with $\left( j_{3},j_{1},j_{2}\right) $ because $\left(
j_{3},j_{1},j_{2}\right) P_{2}^{\prime \prime }\left(
j_{1},j_{3},j_{2}\right) $.}

Let us now consider $R^{\prime }$. Let us consider the allocation $x\left(
i+1,R\right) =\left( j_{1},j_{2},j_{3}\right) $. Since $R^{\prime }$ is a
monotonic transformation of $R$ at $x\left( i+1,R\right) $, it follows that
we can only move to the next element of the ordered set, that is, to $%
x\left( i+2,R\right) $. Note that the top-ranked job for agent 1 at $%
R^{\prime }$ is $j_{1}$. Also, note that the top-ranked job for agent 3 at $%
R^{\prime }$ is $j_{3}$. This implies that only agent 2 can move to $x\left(
i+2,R\right) $, and so $x\left( i+2,R\right) $ must coincide with $\left(
j_{3},j_{1},j_{2}\right) =x\left( i,R\right) $, which contradicts the
assumption that the elements of $F\left( R\right) $ can be ordered as $%
x\left( 1,R\right) ,x\left( 2,R\right) ,x\left( 3,R\right) $. Thus, $F$ does
not satisfy {\it rotation monotonicity}.
\end{example}

Given this, we focus on two classes of job rotation problem that satisfy {\it rotation monotonicity} and thus can be implemented in rotation programs.

\subsection{A Job Rotation Problem With Restricted Domain}

There are situations in which there is a common best/worst job among the
available ones. For instance, suppose that the head of an economics
department needs to allocate one microeconomics course to each of its
microeconomics teachers. Courses can be ranked according to their sizes. The
best possible assignment for everyone is to be assigned to the PhD course
with the lowest number of students, whereas the common worst possible
outcome for every teacher is to be assigned to the largest possible class at
the undergraduate level.

In what follows, we consider situations in which
there is a common best  job, which is denoted by $j_{1}^{\ast }$.
Since situations in which there is a common worst job can be treated
symmetrically, we omit their analysis here. The set of
jobs $J$ is  given by $\left\{ j_{1}^{\ast },j_{2},...,j_{n}\right\} $.
Let $\mathcal{\bar{%
R}}$ be  preference domain such that \\
$ \mathcal{\bar{R}}=\left\{ R\in \mathcal{R}|\text{for all }i\in N\text{, }%
\arg \max_{J}R_{i}=\left\{ j_{1}^{\ast }\right\} \right\}$.
With abuse of notation, we also use $\mathcal{\bar{R}}$ to denote the set of
all (extended) preference profiles.
The next result show that the efficient solution $F$ defined over $\mathcal{%
\bar{R}}$ is implementable in rotation programs.

\begin{theorem}
\label{Th5}
 $F:\mathcal{\bar{R}}\rightarrow \mathcal{\bar{J}}_{0}$
is implementable in rotation programs.
\end{theorem}

The intuition behind this theorem is that for each $R$, elements of $F\left(
R\right) $ can be arranged circularly as $x\left( 1,R\right) ,...,x\left(
m,R\right) ,x\left( 1,R\right) $ such that no two consecutive allocations of
the arrangement allocate $j_{1}^{\ast }$ to the same agent. Thus, the
ordered set required by {\it rotation monotonicity} can be set as $x\left( 1,R\right)
,...,x\left( m,R\right) $. Take any $R^{\prime }$ such that $F\left(
R\right) \neq F\left( R^{\prime }\right) $. Since $F$ is monotonic, it
follows that there exists an $x\left( i,R\right) \in F\left( R\right) $ for
which it holds that
$
x\left( i,R\right) R_{\ell }z\text{ and }zP_{\ell }^{\prime }x\left(
i,R\right)
$
for some agent $\ell \in N$ and an allocation $z\in \bar{J}$. Since, by the
way we arranged the elements of $F\left( R\right) $, it holds that for all $%
k\neq i$, $x\left( k+1,R\right) P_{j}^{\prime }x\left( k,R\right) $ for some
agent $j$, it is clear that $F$ satisfies {\it rotation monotonicity}.

\subsection{A Job Rotation Problem With Partially Informed Planner}

As another application we  consider a  scenario in which the planner
knows that two agents have the same top choice. Specifically, for agent $i$%
's linear ordering $R_{i}\subseteq J\times J$, let $\tau \left( R_{i}\right)
$ denote the top-ranked job of agent $i$ at $R_{i}$. We assume that
planner knows that both agent 1 and agent 2 have a common top-ranked job, although he does not necessarily know which job this is,
and that the domain of admissible profiles of linear orderings is given by $
\mathcal{\hat{R}}=\left\{ R\in \mathcal{R}|\tau \left( R_{1}\right) =\tau
\left( R_{2}\right) \right\}
$. With abuse of notation, we also use $\mathcal{\hat{R}}$ to denote the set of
all (extended) preference profiles over $\bar{J}$.

We are interested in implementing a subsolution $\phi :\mathcal{\hat{R}%
\longrightarrow \bar{J}}_{0}$ of the efficient solution. We construct $\phi $
at $R$ by following three sequential steps: 
\textbf{Step 1:} Assign $\tau \left( R_{1}\right) $ either to agent 1 or to
agent 2.
 \textbf{Step 2:} Assign the remaining jobs $J\backslash \left\{ \tau \left(
R_{1}\right) \right\} $ to $N\backslash \left\{ 1,2\right\} $ in a Pareto
efficient way.
 \textbf{Step 3:} Assign the remaining job to agent 2 if agent 1 has received
his top-ranked job, otherwise, assign it to agent 1.
The set $\phi \left( R\right) $ can be thought of as the set of outcomes
generated by an underlying random {\it serial dictatorship mechanism} \citep{Abdulkadiroglu1998}, in which
the only permutations that are admissible are those in which the first
agent and the last agent of the ordering are respectively either agent 1
and agent 2 or agent 2 and agent 1. Observe that $\#\phi \left( R\right)
=2m$, where $m$ is the number of such  allocations at $R$  where all jobs except $\tau \left( R_{1}\right) $ are assigned to agents $N \setminus {1,2}$ in efficient way (agent 2 getting the leftover). It follows that {\it Property M} is always satisfied by $\phi $ and \Cref{cor4} applies. Thus it suffices to prove that {\it rotation monotonicity} is satisfied. 

Fix any $R\in \mathcal{\hat{R}}$ and any $x\in \phi \left( R\right) $. Let $%
\hat{x}$ be the allocation obtained from $x$ in which the job assigned to
agent 1 under $x$ is assigned to agent 2 under $\hat{x}$, the job assigned
to agent 2 under $x$ is assigned to agent 1 under $\hat{x}$, whereas all
other assignments are unchanged. That is, $\hat{x}_{1}=x_{2}$, $\hat{x}%
_{2}=x_{1}$, and $\hat{x}_{i}=x_{i}$ for every agent $i\neq 1,2$. Observe
that $\hat{x}\in \phi \left( R\right) $ if and only if $x\in \phi \left(
R\right) $.
The next result show that the efficient solution $\phi $ is implementable in rotation programs. This result is obtained by requiring that
the ordered set
\begin{equation*}
\phi \left( R\right) =\left\{ x\left( 1,R\right) ,x\left( 2,R\right)
,...,x\left( 2n-1,R\right) ,x\left( 2m,R\right) \right\}
\end{equation*}
satisfies the following properties for all $i\in \left\{ 1,...,2m\right\} $: (1) If $i$ is odd, then $x_{1}\left( i,R\right) =\tau \left( R_{1}\right) $. (2) If $i$ is even, then $x_{2}\left( i,R\right) =\tau \left( R_{2}\right)
$. (3) If $x\left( i,R\right) =x$ and $i$ is odd, then $x\left( i+1,R\right) =%
\hat{x}$.
$\phi \left( R\right) $ is implementable in rotation programs
because we can devise a rights structure that allows agent 1 (agent 2) to
be effective in moving from the outcome $x\left( i,R\right) $ to $x\left(
i+1,R\right) $ provided that $i$ is even (odd). The reason is that agent 1
(agent 2) has incentive to move from $x\left( i,R\right) $ to his
top-ranked outcome $x\left( i+1,R\right) $ when $i$ is odd (even).
To see that {\it rotation monotonicity} is satisfied, fix any $R^{\prime }$ such that $\phi
\left( R\right) \neq \phi \left( R^{\prime }\right) $. This implies that at
least one allocation $x\left( i,R\right) \in \phi \left( R\right) $ is
Pareto dominated at $R^{\prime }$, that is, there exists an allocation $z$
such that $zR_{j}^{\prime }x\left( i,R\right) $ for each agent $j\in N$ and
$zP_{j}^{\prime }x\left( i,R\right) $ for some agent $j\in N$.
We can proceed according to whether $\tau \left( R_{1}\right) \neq \tau
\left( R_{1}^{\prime }\right) $.
Suppose that $\tau \left( R_{1}\right) \neq \tau \left( R_{1}^{\prime
}\right) $. This implies that $\tau \left( R_{1}\right) =\tau \left(
R_{2}\right) $ has fallen strictly in agent $j=1,2$'s ranking when the
profile moves from $R$ to $R^{\prime }$. This preference reversal both
agent 1 and agent 2 guarantees that {\it rotation monotonicity} is satisfied for every $%
x\left( i,R\right) \in \phi \left( R\right) $.
Suppose that $\tau \left( R_{1}\right) =\tau \left( R_{1}^{\prime
}\right) $. We have already observed that at $R$, it holds that
$
x\left( i+1,R\right) P_{2}x\left( i,R\right)
$
if $i$ is odd, and that
$
x\left( i+1,R\right) P_{1}x\left( i,R\right)
$
if $i$ is even. In other words, there is the following cycle among outcomes
in $\phi \left( R\right) $:
\begin{equation*}
x\left( 1,R\right) P_{1}x\left( 2m,R\right) P_{2}x\left( 2n-1,R\right) \cdot
\cdot \cdot x\left( 3,R\right) P_{1}x\left( 2,R\right) P_{2}x\left(
1,R\right)
\end{equation*}%
Since $\tau \left( R_{j}\right) =\tau \left( R_{j}^{\prime }\right) $ for $%
j=1,2$, it follows that the above cycle also exists at $R^{\prime }$. Since $%
\phi \left( R\right) \neq \phi \left( R^{\prime }\right) $, we already know
that there is at least one allocation $x\left( i,R\right) \in \phi \left(
R\right) $ that is Pareto dominated at $R^{\prime }$. Since $x\left(
i,R\right) $ is efficient at $R$, it follows that $x\left( i,R\right) \in
\phi \left( R\right) $ has strictly fallen in the preference ranking of at
least one agent $j\neq 1,2$ when the profile moves from $R$ to $R^{\prime }$%
. It follows that {\it rotation monotonicity} is satisfied.
We have thus proved the following result.

\begin{theorem}
\label{Th6}$\phi $ $:\mathcal{\hat{R}}\rightarrow \mathcal{\bar{%
J}}_{0}$ is implementable in rotation programs.
\end{theorem}

\section{Discussion}

\subsection{An Envy-Free Mechanism}
\label{envy}

We argued in  \Cref{sectJobRotat} that there are SCRs that satisfy {\it rotation monotonicity}, though it is true that some do not. Here, we discuss the possibility for our theory to achieve fairness in  allocation problems. From this perspective, drawing a lottery is the most common way to solve such problems. If there are two different flavors of ice cream in the freezer, and both children
like the same flavor, parents will suggest drawing a lottery. If several tasks are to be
allocated to people, some more laborious than others, the allocation can be decided by drawing a lottery. When a person dies and leaves tangible assets, heirs
may use a lottery to distribute them. However, anyone who has been part of
these situations knows that there will be a lot of discontent ex-post: children crying,
adults cursing, and heirs never talking to each other again. Nevertheless, the literature on mechanism design has not been able to approach the problem of fairness
in any other way than by drawing a lottery \citep{Hofstee1990, Moulin2001, Budish2013}.
This is so even though experimental evidence suggests drawing a lottery is often
not even considered fair (\citet{Eliaz2014}, \citet{Andreoni2020}).
Given these findings, it would be natural to check whether rotation programs can restore fairness. The answer is yes.   However,  there will be limits. To be concrete, let us consider the Gale-Shapley matching model \citep{GaleShapley1962}. The Gale-Shapley algorithm is an algorithm for finding a solution to the stable matching problem.  Depending on how it is used, it can find either the solution that is optimal for the participants on one side of the matching (\cite{RothVandeVate1990}).  Therefore, this algorithm is neither procedurally nor end-state fair. Indeed, it induces a 
large amount of ex-post envy because the best possible matching for one side of the
market is the worst possible matching for agents on the other side of the
market.\footnote{
This opposition of interests can be observed not only in comparing the
optimal stable matchings but also in comparing any two stable matchings
(\cite{Knuth1976}).}
To at least recover ex-ante fairness, \cite{KlausKlijn2006}
consider two probabilistic matching algorithms that assign to each marriage
market a probability distribution over stable matchings (employment by lotto
and the random order mechanism), and they identify two important properties
that help to differentiate them. However, these algorithms can still induce
a large amount of ex-post envy.

The following example illustrates how our notion of implementation in rotation programs can represent a device to restore ex-post fairness in matching environments.

\begin{example}
\label{ex5}
A \textit{marriage problem} is a quadruplet $\left( M,W,P,\mathcal{M}\right)
$ where $M$ is a finite non-empty set of men, with $m$ as a typical element, $W$ is a finite non-empty set of women, with $w$ as a typical element, $P=\left( P_{i}\right) _{i\in M\cup W}$ is a profile of linear orderings
such that (\textit{i}) every man $m\in M$'s preference ordering is a linear
order $P_{m}$ over the set $W\cup \left\{ m\right\} $ and (\textit{ii})
every woman $w\in W$'s preference ordering is a linear order $P_{w}$ over $%
M\cup \left\{ w\right\} $\footnote{%
A linear ordering $P$ over $X$ is a complete, transitive and anti-symmetric
binary relation over $X$. A binary relation $P$\ over $X$ is anti-symmetric
provided that for all $x,y\in X$, if $xPy$ and $yPx$, then $x=y$.}, and $\mathcal{M}$ is a collection of all matchings, with $\mu $ as a
typical element. $\mu :M\cup W\rightarrow M\cup W$ is a bijective function
matching every agent $i\in M\cup W$ either with a partner of the opposite
sex or with herself. If an agent $i$ is matched with herself, we say that
this $i$ is \textit{single} under $\mu $. Let $\left( M,W,P,\mathcal{M}\right) $ be a marriage problem. Every man $%
m$'s preference ordering $P_{m}$ over $W\cup \left\{ m\right\} $ can be
extended to an ordering over the collection $\mathcal{M}$ in the following
way:%
\[
\mu R_{m} \mu ^{\prime }\Leftrightarrow \text{either }%
\mu \left( m\right) P_{m} \mu ^{\prime }\left( m\right)
\text{ or }\mu \left( m\right) =\mu ^{\prime }\left( m\right) \text{,%
\hspace*{4mm}for every }\mu ,\mu ^{\prime }\in \mathcal{M}\text{.}
\]%
Likewise, this can be done for every woman $w\in W$.
A matching $\mu $ is \textit{individually rational} at $R$ if no agent $i\in
M\cup W$ prefers strictly being single to being matched with the partner
assigned by the matching $\mu $; that is, for every agent $i$, either $\mu
\left( i\right) P_{i}i$ or $\mu \left( i\right) =i$. Furthermore, a matching
$\mu $ is \textit{blocked} at $R$ if there are two agents $m$ and $w$ of the
opposite sex who would each prefer strictly to be matched with the other
rather than with the partner assigned by the matching $\mu $; that is, there
is a pair $\left( m,w\right) $ such that
$
wP_{m}\mu \left( m\right) \text{ and }mP_{w}\mu \left( w\right) \text{.}
$
A matching $\mu $ is \textit{stable} at $R$ if it is individually rational
and unblocked at $R$. A matching $\mu $ is \textit{man-optimal stable} at $R$
if it is the best stable matching from the perspective of all the men; that
is, $m$ is stable at $R$ and for every man $m\in M$, $\mu R_{m} \mu ^{\prime }$ for every other stable matching $\mu ^{\prime }$ at $%
R$. The man-optimal stable matching at $R$ is denoted by $\mu _{M}^{R}$. The
woman-optimal stable matching at $R$ is the best stable matching from the
perspective of all the women and it is denoted by $\mu _{W}^{R}$.

Suppose that the objective is to rotate partners between the man-optimal
stable matching and the woman-optimal stable matching for each profile $R$,
that is, $F\left( R\right) =\left\{ \mu _{M}^{R},\mu _{W}^{R}\right\} $.
Suppose there are three men $M=\{m_{1},m_{2},m_{3}\}$ and three women $%
W=\{w_{1},w_{2},w_{3}\}$. Suppose that $\mathcal{R}=\left\{ R,R^{\prime
}\right\} $ and that agents' preferences at $R$ are as follows:%
\medskip
\begin{center}
\begin{tabular}
[c]{c|c|c|c|c|c|c|c|c|c|c|c}%
\multicolumn{6}{c|}{$R$} & \multicolumn{6}{|c}{$R^{\prime}$}\\\hline
\emph{$m_1$} & \emph{$m_2$} & \emph{$m_3$} & \emph{$w_1$} & \emph{$w_2$} & \emph{$w_3$} & \emph{$m_1$} & \emph{$m_2$} & \emph{$m_3$} & \emph{$w_1$} & \emph{$w_2$} & \emph{$w_3$}\\\hline
$w_2$ & $w_3$ & $w_1$ & $m_1$ & $m_2$ & $m_3$ & $w_2$ & $w_3$ & $w_1$ & $m_2$ & $m_3$ & $m_1$ \\
$w_3$ & $w_1$ & $w_2$ & $m_3$ & $m_1$ & $m_2$ & $w_3$ & $w_1$ & $w_2$ & $m_3$ & $m_1$ & $m_2$ \\
$w_1$ & $w_2$ & $w_3$ & $m_2$ & $m_3$ & $m_1$ & $w_1$ & $w_2$ & $w_3$ & $m_1$ & $m_2$ & $m_3$ \\
$m_1$ & $m_2$ & $m_3$ & $w_1$ & $w_2$ & $w_3$ & $m_1$ & $m_2$ & $m_3$ & $w_1$ & $w_2$ & $w_3$ 
\end{tabular}
\end{center}

\medskip

Note that $R_{m}=R_{m}^{\prime }$ for all $m\in M$.
The man-optimal stable matching and the woman-optimal stable matching at $R$
are: $\mu _{M}^{R}(m_1)=w_2$, $\mu _{M}^{R}(m_2)=w_3$, $\mu _{M}^{R}(m_3)=w_1$, $\mu _{W}^{R}(m_1)=w_1$, $\mu _{W}^{R}(m_2)=w_2$ and $\mu _{W}^{R}(m_3)=w_3$.

\noindent Whereas at $R^{\prime }$ they are: $\mu _{M}^{R}(m_1)=w_2$, $\mu _{M}^{R}(m_2)=w_3$, $\mu _{M}^{R}(m_3)=w_1$, $\mu _{W}^{R}(m_1)=w_3$, $\mu _{W}^{R}(m_2)=w_1$ and $\mu _{W}^{R}(m_3)=w_2$,
where $\mu _{M}^{R}=\mu _{M}^{R^{\prime }}$ has $m_{1}$ married to $w_{2}$, $%
m_{2}$ married to $w_{3}$ and $m_{3}$ married to $w_{1}$. It follows that $%
F\left( R\right) =\left\{ \mu _{M}^{R},\mu _{W}^{R}\right\} $ and $F\left(
R^{\prime }\right) =\left\{ \mu _{M}^{R^{\prime }},\mu _{W}^{R^{\prime
}}\right\} $, so that $F\left( R\right) \neq F\left( R^{\prime }\right) $,
and $\#F\left( R^{\prime }\right) >1$. In what follows we show  that $F$
satisfies {\it rotation monotonicity}. Fix $R\in {\cal R}$ and let us consider the order of states $x(1,R)=x(\mu _{W}^{R},R),x(\mu _{M}^{R},R)=x(2,R)$.
Then for every $w\in W$ it holds that $x(\mu _{M}^{R},R)P'_w x(\mu _{W}^{R},R)$ and $x(\mu _{W}^{R},R)R_w x(\mu _{M}^{R},R)$, thus {\it rotation monotonicity} is satisfied w.r.t. $R$.
Finally, fix $R'\in {\cal R}$ and  consider the order of states $x(1,R')=x(\mu _{W}^{R'},R'),x(\mu _{M}^{R'},R')=x(2,R')$. For every $w\in W$ it holds that $x(\mu _{M}^{R'},R')P_w x(\mu _{W}^{R'},R')$ and $x(\mu _{W}^{R'},R')R'_w x(\mu _{M}^{R'},R')$, thus {\it rotation monotonicity} is also satisfied w.r.t. $R'$.

\end{example}

However,  this situation is  unattainable in general because the SCR that picks only the man-optimal and woman-optimal stable matchings can violates 
{\it rotation monotonicity}.\footnote{See Example 6 in \cite{Korpela2021}. } We believe that the identification and the  characterization of classes of   allocation problems that  can be implemented in (a form of) rotation programs is a fruitful area for future research.

\subsection{Concluding Remarks}

This paper studies rotation programs in an implementation framework. A rotation program is an MSS \citep{DemuynckHeringsSaulleSeel2019a} in which states are arranged circularly.
We identify conditions for implementation in MSS of Pareto efficient SCRs by a finite rights structure \citep{KorayYildiz2019}. Implementation in MSS is robust in the following sense: at any preference profile, every non-stable allocation converges to a stable allocation via a sequence of myopic deviations. Moreover, implementation in MSS encompasses implementation in absorbing sets and in generalized stable sets.
 
We identify a sufficient condition for implementation in MSS, named  {\it indirect monotonicity}. This condition is weaker than (Maskin) monotonicity. Furthermore, we show that {\it rotation monotonicity}, when combined with an auxiliary condition, is sufficient for implementation in rotation programs. Rotation monotonicity is necessary and sufficient for implementation when the SCR never selects a single outcome. Finally, we study some  welfare implications of this characterization result.

\section*{\hypertarget{appendixA}{Appendix A}}
\addcontentsline{toc}{section}{Appendix A}

\subsection*{Convergence in  Exchange Economy}
\addcontentsline{toc}{subsection}{Convergence in  Exchange Economy}
 Let us consider the class of exchange economies studied by
\citet{BalbuzanovKotowski2019} and consider the notion of {\it direct exclusion core}. We
show, by means of an example, that free exchange of goods do not necessary
converge to the direct exclusion core. However, the direct exclusion core is
implementable in MSS via a finite rights structure. This implies that irrespective of the initial allocation of objects, it is possible to converge to a direct exclusion core allocation in a finite sequence of coalitional moves.

An {\it economy} is a quadruplet $\left( N,H,P,\omega \right) $ where $N=\left\{ 1,...,n\right\} $ is a finite non-empty set of agents, $H=\{h_{1},...,h_{m}\}$ is a finite set of indivisible objects, called
houses, that can be allocated among the agents, $P=\left( P_{i}\right) _{i\in N}$ is a profile of linear orderings,
where each linear ordering is defined over $H\cup \left\{ h_{0}\right\} $, and the endowment system $\omega :2^{N}\longrightarrow 2^{H}$ is a function that specifies the
houses owned by each coalition.
For each coalition $K\in \mathcal{N}_{0}$, we write  $\omega \left( K\right) =\bigcup
_{T\in \mathcal{K}_{0}}\omega \left( T\right) $. Let us assume that the
endowment system $\omega $ satisfies the following four properties: (A1) \emph{Agency}: $\omega \left( \emptyset \right) =\emptyset $, (A2) \emph{Monotonicity}: $K\subseteq K^{\prime }\implies \omega \left(
K\right) \subseteq \omega \left( K^{\prime }\right) $, (A3) \emph{Exhaustivity}: $\omega (N)=H$, and (A4) \emph{Non-contestability}: For each $h\in H$, there exists $K^{h}\in
\mathcal{N}_{0}$ such that $h\in \omega \left( K\right) \iff K^{h}\subseteq
K $.

Property A1 restricts ownership to agents or groups. Property A2 requires
that a coalition has in its endowment anything that belongs to any
sub-coalition. Property A3 states that the grand coalition $N$ jointly owns
everything. In property A4, coalition $K^{h}$ is called the minimal
controlling coalition of house $h$. It guarantees that each house has a set
of one or more \textquotedblleft co-owners\textquotedblright\ without
opposing and mutually exclusive claims. As \citet[Lemma 1]{BalbuzanovKotowski2019} show, these properties are needed to assure that the direct exclusion core is nonempty.

We assume that each agent may live in at most one house and each house $h\in
H$ may accommodate at most one agent. A house may be vacant and an agent can
be homeless. We can model this latter outcome by the agent's assignment to
an outside option $h_{0}\notin H$, which has unlimited capacity.

An allocation $\mu :N\longrightarrow H\cup \left\{ h_{0}\right\} $ is an
assignment of agents to houses such that $\#\mu ^{-1}\left( h\right) \leq 1$
for all $h\in H$. We write $%
\mu
(K)$ to denote $\bigcup _{i\in K}\mu \left( i\right) $ for any $K\in \mathcal{N}%
_{0}$.
Let $\left( N,H,R,\omega \right) $ be an economy. Every linear ordering $%
R_{i}$ can be extended to an ordering over the collection $\mathcal{M}$ of
allocations in the following way: 
$
\mu R_{i}\mu ^{\prime }\iff \text{ either }\mu \left( i\right) P_{i}\mu
^{\prime }\left( i\right) \text{ or }\mu \left( i\right) =\mu ^{\prime
}\left( i\right) \text{,}
$
for all $\mu ,\mu ^{\prime }\in \mathcal{M}$. With little abuse of notation, we
denote both by $R_{i}$. Let $\mathcal{R}$ denote the class of admissible
preference profiles of extended preferences.

\begin{definition}
\label{def8}
Given an economy $\left( N,H,R,\omega \right) $, a coalition $K\in \mathcal{N%
}_{0}$ can {\it directly exclusion block} the allocation $
\mu$ at $R$ with allocation $\sigma $ if 

\noindent (a) $\sigma (i)P_{i}\mu \left(
i\right) $ for all $i\in K$ and 

\noindent (b)  $\mu(j)P_{j}\sigma (j)\implies \mu
(j)\in \omega (K)$ for all $j\in N\backslash K$.
\end{definition}

To speak, a coalition can directly exclusion block an assignment whenever each member strictly gains from
an alternative and anyone harmed by the reallocation is excluded from a
house belonging to the coalition. The \textit{direct exclusion core} is the
set of allocations that cannot be directly exclusion blocked by any nonempty
coalition.

\begin{definition}[{\it Direct Exclusion core}]
Given an economy $\left( N,H,R,\omega \right) $, its \textit{direct
exclusion core}, denoted by $CO\left( R,\omega \right) $, is defined by 
$
CO\left( R,\omega \right) =\{ \mu \in \mathcal{M}|\text{no coalition}$ \newline$
\text{can directly exclusion block }\mu \text{ at }R\} \text{.}
$
\end{definition}

Thus, no coalition can gainfully destabilize a direct exclusion core
allocation by invoking their collective exclusion rights. \citet[Lemma 1]{BalbuzanovKotowski2019} show that the direct exclusion core is never empty
and all its allocations are Pareto efficient.

Let us show that the direct exclusion core does not satisfy any external
stability requirement. To this end, let us represent an allocation $\mu $ by
a permutation matrix with columns indexed by elements of $N$ and rows
indexed by elements of $H\cup \left\{ h_{0}\right\} $, where $h_{0}$ is the
last row. If for some $h\in H\cup \left\{ h_{0}\right\} $ and some $i\in N$,
entry $\mu _{hi}=1$, then good $h$ has been assigned to agent $i$.

Let us consider an economy with three agents and three houses.\footnote{We borrow this example from \citet[pp.12-13]{DemuynckHeringsSaulleSeel2019b}.} Each house $%
i\in H$ is owned by agent $i$ and agents' preferences are given in the table below.%
\begin{figure}[ht]
    \centering
\begin{minipage}{0.3\textwidth}
\begin{tabular}{lll}
& $R$ &  \\ \hline
1 & \multicolumn{1}{|l}{2} & \multicolumn{1}{|l}{3} \\ \hline
$2$ & \multicolumn{1}{|l}{$3$} & \multicolumn{1}{|l}{$1$} \\
$3$ & \multicolumn{1}{|l}{$1$} & \multicolumn{1}{|l}{$2$} \\
$1$ & \multicolumn{1}{|l}{$2$} & \multicolumn{1}{|l}{$3$} \\
$h_{0}$ & \multicolumn{1}{|l}{$h_{0}$} & \multicolumn{1}{|l}{$h_{0}$}%
\end{tabular}%
\text{}
\end{minipage}
\begin{minipage}{0.3\textwidth}
$\mu =\left[
\begin{array}{ccc}
0 & 0 & 1 \\
1 & 0 & 0 \\
0 & 1 & 0 \\
0 & 0 & 0%
\end{array}%
\right] \text{}$
\end{minipage}

\end{figure}
It can be checked that the direct exclusion core at $R$ consist of the allocation $%
\mu $.
Let us consider the following allocations:

\begin{center}
$\sigma ^{1}=\left[
\begin{array}{ccc}
0 & 1 & 0 \\
1 & 0 & 0 \\
0 & 0 & 1 \\
0 & 0 & 0%
\end{array}%
\right] \text{, }\sigma ^{2}=\left[
\begin{array}{ccc}
1 & 0 & 0 \\
0 & 0 & 1 \\
0 & 1 & 0 \\
0 & 0 & 0%
\end{array}%
\right] \text{ and }\sigma ^{3}=\left[
\begin{array}{ccc}
0 & 0 & 1 \\
0 & 1 & 0 \\
1 & 0 & 0 \\
0 & 0 & 0%
\end{array}%
\right] \text{.}$
\end{center}%

Although the direct exclusion core is not empty, the process of `free'
exchange of houses may not lead to $\mu $ because such a process may cycle.
Indeed, agents may myopically cycle around $\sigma ^{1}$, $\sigma ^{2}$ and $%
\sigma ^{3}$.

To see it, note that for each agent $i$, his endowment $\omega \left(
i\right) =i$ corresponds to his third choice--his last choice is to become
homeless. Therefore, given this initial situation, coalition $\left\{
1,2\right\} $ can trade so that they can achieve the allocation $\sigma ^{1}$%
. At $\sigma ^{1}$, agent 1 obtains his first best choice. Thus, coalition $%
\left\{ 2,3\right\} $ is the only coalition that can achieve a strict
improvement. The only allocation that $\left\{ 2,3\right\} $ can move to is
allocation $\sigma ^{2}$, where agent 2 obtains is first best choice. At $%
\sigma ^{2}$, only coalition $\left\{ 1,3\right\} $ can achieve a strict
improvement by moving to the only attainable allocation $\sigma ^{3}$, where
agent 3 obtains is first best choice. At $\sigma ^{3}$, only coalition $%
\left\{ 1,2\right\} $ can achieve a strict improvement by moving to the only
attainable allocation $\sigma ^{1}$. Therefore, free exchange may lock agents in a cycle of exchanges.

A natural question that arises from the preceding example is whether it is
possible to achieve the direct exclusion core by means of a different exchange
process. The answer is provided by \Cref{cor2}, which shows that the direct exclusion core is
implementable in MSS via a finite rights structure. To formalize our answer,
fix any endowment system $\omega $ satisfying the above four properties. Let
us define $F_{\omega }^{CO}$ by $F_{\omega }^{CO}\left( R\right) =CO\left(
R,\omega \right) $ for all $R\in \mathcal{R}$.

\begin{corollary}
\label{cor2}
Fix any endowment system $\omega $ satisfying properties A1-A4. $F_{\omega
}^{CO}$ is implementable in MSS via a finite rights structure.
\end{corollary}

\subsection*{Convergence In Matching}
\addcontentsline{toc}{subsection}{Convergence in  Matching}
\noindent As a second application, we consider a two-sided, one-to-one matching model, namely the ``marriage problem".  
A marriage problem is a market without transfers where the sides of the market are, for example, workers and firms (job matching), medical students and hospitals (matching of students to internships), students and advisors (matching of students to thesis advisors). The two sided of the markets are simply referred as ``men" and ``women", hence the name ``marriage problem".
An output of the model is termed a matching, which pairs each woman with at
most one man, and each man with at most one woman. Roughly speaking, a matching
is stable when there is no blocking pair, that is, no pair of agents are
better off with each other than with their assigned partners. A formal description of this matching model is  presented in \Cref{envy}.
There are two prominent models describing the marriage problem: the Gale-Shapley model \citep{GaleShapley1962} and the Knut model  \citep{Knuth1976}.
The former studies stability for marriage problems
allowing the possibly for agents to be single. The latter is a pure matching model in which no agents is allowed to be single (and thus the number of men and women is assumed to be the same).   \citet{RothVandeVate1990} show that, the set of stable matching in the  Gale-Shapley model exhibits a convergence property, that is,  for any non stable matching there exist a myopic improvement path  to a  stable matching. On the contrary, for the Knut model, no general convergence result is provided. Moreover \citet{Tamura1993} shows that, under usual matching rules, when there are at least four women, there exists  preferences such that agents  cycle among non stable matchings. Our  next result fills the gap. Indeed,  since  a stable matching in the marriage problem is monotonic and efficient, we establish, as a  corollary to \Cref{Th1},  that the set of stable matching in the Knut model is implementable in MSS and 
thus there exists a mechanism such that a  converge property  in the Knut model is restored.\footnote{The proof of \Cref{cor3} is omitted.}

\begin{corollary}
\label{cor3}
The set of stable matching in the Knut model is implementable in MSS via a finite right structure.
\end{corollary}

Note that,  under  usual matching rules, \citet{DemuynckHeringsSaulleSeel2019a} show that the MSS is a superset of the set of stable matchings. From this point of view, \Cref{cor3}  further enlighten the relation   between the MSS and the set of stable matchings.
Moreover, it  suggests  that the  implementation in right structure could represent a tool to refine the MSS whenever its prediction under canonical rules is too loose.  Since  this  conjecture  overcomes  the purposes  of  the present work, we leave it as an avenue for future research.

\section*{\hypertarget{appendix}{Appendix B}}
\addcontentsline{toc}{section}{Appendix B}

\subsection*{Proofs}
\addcontentsline{toc}{subsection}{Proofs}

\noindent{\bf Proof of \Cref{Th1}.}
The state space $S$ consists of $S=Gr(F) \cup Z$. Since $Z$ finite, it follows that $S$ is finite as well. The  outcome function $h$ is defined such that $h(z,R)=z$ for all $(z,R) \in S$ and $h(z)=z$ for all $z \in Z$.
The code of rights $\gamma$  is given by  the following five rules:
\medskip

 \noindent  {\textbf{RULE 1:}} $\{i\} \in \gamma((z,R),(x,R))$ for all $R \in \mathcal{R}$, all $z,x \in F(R)$, and all $i \in N$,
\medskip

 \noindent  {\textbf{RULE 2:}} $\{i\} \in \gamma((z,R),x)$ if $x \in L_{i}(z,R)$,
\medskip

 \noindent  {\textbf{RULE 3:}} $\{i\} \in \gamma(x,(z,R))$ for all $x,(z,R) \in S$, and all $i \in N$, 
\medskip

 \noindent  {\textbf{RULE 4:}} $\{i\} \in \gamma(x,y)$ for all $x,y \in S$, and all $i \in N$, and
\medskip

 \noindent  {\textbf{RULE 5:}} $\gamma(s,s')=\emptyset$ for any other $s, s' \in S$.
\medskip

Let us show that the rights structure $\Gamma=(S,h,\gamma)$ defined above implements $F$ in MSS if $F$ is  efficient and indirect monotonic. To this end, suppose that $F$ is  efficient and indirect monotonic. The following lemmata will be useful in proving our result. 
To proceed with our lemmata, we need the following additional definitions. For each $R,R' \in \mathcal{R}$:

\begin{align*}
\hspace{-1.0cm}
M(R)\equiv\{(z,R) \mid z \in F(R)\} \subseteq S & \ \ \ \ \ U(R)\equiv\{ z \in Z \mid Z\subseteq L_{i}(z,R) \; \textrm{for all} \; i \in N \};
\end{align*}
\begin{align*}
\hspace{-1.0cm}Q\left( R,R^{\prime }\right) \equiv\left\{ 
\begin{tabular}{l|l}
$\left( z^{\prime },R^{\prime }\right)\in  M\left( R^{\prime
}\right)$ & there does not exist any myopic
improvement \\ 
& path from $\left( z^{\prime },R^{\prime }\right) $ to $M\left( R\right) \cup U\left( R\right) $ at $R$%
\end{tabular}%
\right\};
\end{align*}
\begin{align*}
\hspace{-1.0cm}Q(R)\equiv\underset{R' \in \mathcal{R}}{\bigcup} Q(R,R'). 
\end{align*}
Since $S$ is finite, the property of asymptotic external stability of  \Cref{def5} is equivalent to the property of iterated external stability, which is defined in a footnote of \Cref{sect untermediate step}. 
Fix any profile $R$. The objective of the following lemmata is to show that 
\begin{align*}
\hspace{-1cm}MSS(\mathnormal{\Gamma},R)=M(R) \cup U(R) \cup Q(R) \ \ \text{and} \ \  F(R)=h\circ (M(R) \cup U(R) \cup Q(R)). 
\end{align*}

\begin{lemma}
\label{lem1}
There is a finite myopic improvement path to $M(R) \cup U(R)$ at $R$ from every state $s \in Z \setminus U(R)$.
\end{lemma}

\noindent {\bf Proof of  \Cref{lem1}.} 
Take any $s \in Z\setminus U(R)$. If $U(R)\neq \emptyset$, then there exists  a one step myopic improvement path from $s$ to $U(R)$, by Rule 4. Otherwise, suppose that $U(R) = \emptyset$. We divide the rest of the proof in two parts according to whether $s \notin F(R)$ or not.
\medskip

\noindent {\bf Case 1}:  $s \notin F(R)$.
Suppose that $sR_ih(s')$ for all $i \in N$ and all $s'\in M(R)$. Since $s'\in M(R)$ and $F$ satisfies efficiency, it holds that $sI_ih(s')$ for all $i \in N$. Since $R \in \mathcal{R}$, it follows that $s=h(s')$, and so  $s\in F(R)$, which is  a contradiction. 
Therefore, it must be the case that there exists an $s'\in M(R)$ such that $h(s')P_is$ for some $i\in N$. Hence, by Rule 3,  there exists a one-step improvement path from $s$ to $M(R)$ at $R$. 
\medskip

 \noindent {\bf Case 2}: $s \in F(R)$. 
Suppose that there exists an agent $i \in N$ such that $h(s')P_is$ for some $s'\in M(R)$. By Rule 3, 
 there exists a one step myopic improvement path from $s$ to $M(R)$ at $R$. Otherwise,  suppose that $sR_ih(s')$ for all $s' \in M(R)$ and for all $i\in N$.  Efficiency of $F$ implies that $h(s')I_{N}s$ for all $s' \in M(R)$, and so $h(s')=s$ because $R \in \mathcal{R}$. However, since $U(R) = \emptyset$, there exists $s'' \in Z$ and an agent $i \in N$ such that $s''P_{i}s$. Note that agent $i$ has the power to move from $s$ to $s'$ by Rule 4 and the incentive to do so since $s''P_{i}s$. 
 Since $F$ satisfies  efficiency and $s\in F(R)$, there must exist another agent $j \in N\setminus\{i\}$ such that $sP_{j}s''$. Since $s\in F(R)$, by assumption, it follows that $(s,R)\in M(R)$. By Rule 3, agent $j$ can move from $s''$ to $(s,R)$. Hence, we have established  a two-step myopic improvement path at $R$ from $s$ to $(s,R) \in M(R)$---that is, $i\in \gamma(s,s'')$ and $s''P_is$ and $j\in \gamma(s'',(s,R))$ and $h(s,R)P_js''$.  $\hfill \blacksquare$

\begin{lemma}
\label{lem2}
For any $R' \in \mathcal{R}$, the set $Q(R,R')$ satisfies deterrence of external deviations and $h\left(Q(R,R')\right)=\{h(s)\in Z|s\in Q(R,R')\} \subseteq F(R)$. 
\end{lemma}

\noindent {\bf Proof of \Cref{lem2}.} 
Suppose that $Q(R,R') \ne \emptyset$ for some $R' \in \mathcal{R}$. Otherwise, there is nothing to be proved.  
Let us first prove that $h\left(Q(R,R')\right) \subseteq F(R)$. By definition,  $Q(R,R') \subseteq M(R')$. Take any $(z',R') \in Q(R,R')$. Assume, to the contrary, that  $h(z',R')=z' \notin F(R)$. 
Suppose that there exists an agent $i \in N$ such that $yP_iz'$ for some $y\in L_i(z',R')$. Then, by Rule 2, agent $i\in \gamma((z',R'),y)$ since $y\in L_i(z',R')$.
An immediate contradiction is obtained if $y\in U(R)$ because there is a one step  myopic improvement from $Q(R,R')$ to $U(R)$. Suppose $y\in Z\setminus U(R)$. By \Cref{lem1}, there is a  finite myopic improvement path from $y$ to $M(R)\cup U(R)$ . Therefore, there exists a finite myopic improvement path from $(z',R')$ to $M(R)\cup U(R)$, which contradicts the definition of $Q(R,R')$. 
Thus, it has to be that  $L_i(z',R') \subseteq L_i(z',R)$ for all $i \in N$. 

Let us proceed according to whether $\{z\}=F(R')$ or not. Suppose that $\{z\}=F(R')$. Since $F$ satisfies {\it indirect monotonicity} and $L_i(z',R') \subseteq L_i(z',R)$ for all $i \in N$, it must be the case that $z \in F(R)$, which is a contradiction. 
Suppose that $\{z\}\neq F(R')$. Since  $z'\in F(R')\setminus F(R)$ and since  $L_i(z',R') \subseteq L_i(z',R)$ for all $i \in N$, {\it indirect monotonicity} implies that there exist a  sequence of outcomes  $\{z_{1}\ldots,z_{h}\}\subseteq F(R')$ with $z'=z_{1}$ and $z\neq z_h$  a sequence of agents $i_{1},\ldots,i_{h-1}$ such that 
 (i) $z_{k+1}  P_{i_{k}} z_{k}$ \ for all $k \in \{1, \ldots,h-1\}$ 
and
 (ii) $L_i(z_h,R')\not\subseteq L_i(z_h,R)$ for some $i \in N$.

By Rule 1, part (i) of {\it indirect monotonicity} implies that there exists a finite myopic improvement path from $(z',R')$ to $(z_h, R')\in M(R')$ at $R$. Part (ii) of {\it indirect monotonicity} implies that   there exists a state $y\in L_i(z_h,R')$ such that $yP_iz_h$. By Rule 2, $\{i\}\in \gamma((z_h,R'),y)$.
An immediate contradiction is obtained whenever $y\in U(R)$ because there is a finite myopic improvement path from $(z',R')$ to $U(R)$ at $R$. 
Suppose that  $y\in Z\backslash U(R)$. Then, by \Cref{lem1}, there exists a finite myopic improvement path from  $y$ to $M(R)\cup U(R)$ at $R$. Therefore, there exists a finite myopic improvement path from $(z',R')$ to $M(R) \cup U(R)$ at $R$, which contradicts our initial supposition that $(z',R') \in Q(R,R')$. We conclude that  $h(Q(R,R')) \subseteq F(R)$. 

To complete the proof of \Cref{lem2},  let us show  that $Q(R,R') \subseteq M(R')$ satisfies deterrence of external deviations at $R$. The only way to get out of this set is to use either Rule 1 or Rule 2. Therefore, from any state of  $Q(R,R')$, agents can only deviate to $M(R') \setminus Q(R,R')$ or $Z$. Note that if  $M(R') \setminus Q(R,R')\neq \emptyset$, then there exists a  myopic improvement path to $M(R) \cup U(R)$ at $R$, by the definition of  $Q(R,R')$. Also, note that from any state in $Z \setminus U(R)$, there exists a finite myopic improvement path to $M(R) \cup U(R)$ at $R$, by \Cref{lem1}. Hence, if an agent could benefit by deviating from a state $s \in Q(R,R')$ to a state outside of $Q(R,R')$ at $R$, there would exist a  myopic improvement path from $s$ to $M(R) \cup U(R)$ at $R$, which would contradict the definition of $Q(R,R')$. $\hfill \blacksquare$

\begin{lemma}
\label{lem3}
If $V$ is a nonempty subset of $S$ satisfying both deterrence of external deviations and iterated external stability at $(\Gamma,R)$, then $M(R)\subseteq V$.
\end{lemma}

\noindent {\bf Proof of  \Cref{lem3}.} 
Let $V$ be a nonempty subset of $S$ satisfying both deterrence of external deviations and iterated external stability at $(\Gamma,R)$. We show that $M(R)\subseteq V$. We proceed in two steps.
\medskip

\noindent {\bf Step 1}:  $M(R)\cap V\neq \emptyset$. 
For the sake of contradiction, let $M(R)\cap V=\emptyset$. Then, by iterated external stability of $V$,  there exists a  sequence of states $s_{1}, \ldots,s_{m}$ with $s_1\in M(R)$  and a collection of coalitions $K_{1}, \ldots,K_{m-1}$ such that, for  $j=1, \ldots,m-1$, $K_{j} \in \gamma(s_{j},s_{j+1})$  and  $h(s_{j+1})P_{K_{j}}h(s_{j})$. Moreover, $s_m\in V$.   
By definition of $\gamma$, by the fact that $s_1\in M(R)$ and that $h(s_{j+1})P_{K_{j}}h(s_{j})$, we have that only Rule 1 applies, and so it has to be that $\{s_1,...,s_m\}\subseteq M(R)$. Therefore, $s_m\in M(R)\cap V$, which is a contradiction.
 \medskip

\noindent  {\bf Step 2:}  $M(R)\subseteq  V$. 
Take any $s\in M(R)$. Assume, to the contrary, that $s\notin V$. Since, by Step 1, $M(R)\cap V\neq \emptyset$, take any $s'\in M(R)\cap V$. Since $s,s' \in M(R)$, it must be the case that $h(s) \neq h(s')$. Suppose that for some $i\in N$, $h(s)P_ih(s')$. By Rule 1, agent $i$ can move from $s'$ to $s$, which contradicts the property of  deterrence of external deviations of $V$.  Therefore, it has to be that $h(s')R_Nh(s)$. Since $R \in \mathcal{R}$ and $h(s) \neq h(s')$, it follows that $h(s')P_ih(s)$ for some $i \in N$. Since $F$ is efficient, it follows that $h(s) \notin F(R)$, and so $s \notin M(R)$, which is a contradiction. Since the choice of $s'$ is arbitrary and since, moreover, $s\in M(R)$, it follows that $M(R)\cap V= \emptyset$, which is a contradiction. Thus, it has to be that $M(R)\subseteq V$.
 $\hfill \blacksquare$

\begin{lemma}
\label{lem4}
The set $M(R)\cup U(R) \cup Q(R)$ satisfies both deterrence of external deviations and iterated external stability at $(\Gamma,R)$. Moreover, $F\left(  R\right)  =h\circ (M(R) \cup U(R) \cup Q(R))$.
\end{lemma}

\noindent {\bf Proof of  \Cref{lem4}.} 
By definition of  $\Gamma$, the set $M(R)$ satisfies  deterrence of external deviations. By \Cref{lem2}, the set $Q(R)$ satisfies deterrence of external deviations. By definition, the set  $U(R)$  satisfies deterrence of external deviations. Deterrence of external deviations is therefore satisfied by $M(R) \cup U(R) \cup Q(R)$.
 By \Cref{lem1}, there is a finite myopic improvement path from   $Z \setminus U(R)$ to $M(R) \cup U(R)$ at $R$. 
For any $R' \in \mathcal{R}$, by the definition of $Q(R,R')$, there is a myopic improvement path from  $M(R') \setminus Q(R,R')$ to $M(R) \cup U(R)$ at $R$. This implies that   for any  state outside of $M(R) \cup U(R) \cup Q(R)$ there is a myopic improvement path to $M(R) \cup U(R)$ at $R$, and so iterated external stability is satisfied by $M(R) \cup U(R) \cup Q(R)$. 
 $\hfill \blacksquare$

\begin{lemma}
\label{lem5}
If $V$ is a nonempty subset of $S$ satisfying both deterrence of external deviations and iterated external stability at $(\Gamma,R)$, then $M(R)\cup U(R) \cup Q(R)\subseteq V$. 
\end{lemma}

\noindent {\bf Proof of  \Cref{lem5}.}  
By \Cref{lem3}, we already know that $M\left( R\right) \subseteq V$. By
iterated external stability of $V$, it has to be that $U(R)\subseteq V$%
---the reason is that no myopic improvement path can begin from a
unanimously best outcome. We are left to show that $Q(R)\subseteq V$. To
this end, take any $R^{\prime }\in \mathcal{R}$. Since $Q(R,R^{\prime })$
satisfies deterrence of external deviations at $\left( \Gamma ,R\right) $ by
\Cref{lem2}, it follows that $Q(R,R^{\prime })\subseteq V$, otherwise,
iterated external stability of $V$ is violated by the fact that $%
Q(R,R^{\prime })$ satisfies deterrence of external deviations. Since $%
R^{\prime }$ is arbitrary, we conclude that $Q(R)\subseteq V$. Thus, $%
M(R)\cup U(R)\cup Q(R)\subseteq V$. $\hfill \blacksquare$

\begin{lemma}
\label{lem6}
$M(R)\cup U(R) \cup Q(R)=MSS(\Gamma,R)$ 
\end{lemma}

\noindent {\bf Proof of  \Cref{lem6}.} 
\Cref{lem4} implies that the set $M(R)\cup U(R) \cup Q(R)$ satisfies both deterrence of external deviations and iterated external stability at $(\Gamma,R)$. \Cref{lem5} implies that the set $M(R)\cup U(R) \cup Q(R)$ is the smallest nonempty set satisfying these two properties. Therefore, the unique MSS of $(\Gamma, R)$ consists  of $M(R)\cup U(R) \cup Q(R)$.
$\hfill \blacksquare$

\begin{lemma}
\label{lem7}
 $F\left(  R\right)  =h\circ (M(R) \cup U(R) \cup Q(R))$.
\end{lemma}

\noindent {\bf Proof of \Cref{lem7}.} Let us show that $F\left(  R\right)  =h\circ M(R) \cup U(R) \cup Q(R) $. Clearly, $F(R) \subseteq h\circ M(R)$, and so $F\left(  R\right)  \subseteq h\circ M(R) \cup U(R) \cup Q(R)  $. For the converse, \Cref{lem2} implies that $h\circ Q(R,R') \subseteq F(R)$ for all $R' \in \mathcal{R}$. Since $F$ is efficient, it follows that $U(R) \subseteq F(R)$. Moreover, by definition of $M(R)$, it follows that $h\circ M(R) \subseteq F(R)$. Therefore, $F\left(  R \right)=h\circ M(R) \cup U(R) \cup Q(R) $. $\hfill \blacksquare$
\medskip

\noindent {\textbf {Proof of   \Cref{cor2}}.}
Fix any endowment system $\omega $ satisfying properties A1-A4. $F_{\omega
}^{CO}$ is Pareto efficient because the direct exclusion core is efficient. In
light of \Cref{cor1}, we need only to show that $F_{\omega }^{CO}$ is
monotonic. To this end, take any $\mu \in F_{\omega }^{CO}\left( R\right) $
for some $R\in \mathcal{R}$. Take any $R^{\prime }\in \mathcal{R}$ such that
$L_{i}\left( \mu ,R\right) \subseteq L_{i}\left( \mu ,R^{\prime }\right) $
for all $i$. Let us show that $\mu \in F_{\omega }^{CO}\left( R^{\prime
}\right) =CO\left( R^{\prime },\omega \right) $. Since $\mu \in CO\left(
R,\omega \right) $, it follows that no coalition can directly exclusion
block $\mu $ at $R$. That is, for all $K\in \mathcal{N}_{0}$ and for all $%
\sigma \in \mathcal{M}$, $\mu \left( i\right) R_{i}\sigma \left( i\right) $
for some $i\in K$ or [$\mu \left( j\right) P_{j}\sigma \left( j\right) $ for
some $j\in N\backslash K$ and $\mu \left( j\right) \notin \omega \left(
K\right) $]. If $\mu \left( i\right) R_{i}\sigma \left( i\right) $ for some $%
i\in K$, it follows from the fact that $R^{\prime }$ is a monotonic
transformation of $R$ at $\mu $ that $\mu \left( i\right) R_{i}^{\prime
}\sigma \left( i\right) $ for some $i\in K$. If $\mu \left( j\right)
P_{j}\sigma \left( j\right) $ for some $j\in N\backslash K$ and $\mu \left(
j\right) \notin \omega \left( K\right) $, it follows from the the fact that $%
R^{\prime }$ is a monotonic transformation of $R$ at $\mu $ and the fact
that $R_{j}$ is a linear ordering that $\mu \left( j\right) P_{j}^{\prime
}\sigma \left( j\right) $ for some $j\in N\backslash K$ and $\mu \left(
j\right) \notin \omega \left( K\right) $. We have that no coalition can
directly exclusion block $\mu $ at $R^{\prime }$. Thus, $F_{\omega }^{CO}$
is monotonic.$\hfill \blacksquare$
\medskip

\noindent \textbf{Proof of   \Cref{Th3}.} Suppose that $%
\Gamma $ implements $F$ in rotation program. Fix any $R$. Then, the set $%
MSS\left( \Gamma ,R\right) $ is partitioned in rotation programs $\left\{
S_{1},...,S_{m}\right\} $ such that $h\circ S_{i}=F\left( R\right) $ for all
$i=1,...,J$. Fix any rotation program $S_{j}=\{s_1,...,s_m\}$ for some $m\in\mathbb{N}$. Let $x\left( i,R\right)=
s_i=h\left( s_{i}\right) $ for all $s_{i}\in S_{j}$. Thus, $F\left( R\right) $
is an ordered set of $\#S_{j}=m\geq 1$ outcomes. Fix any $R^{\prime }$ such
that $F\left( R^{\prime }\right) \neq F\left( R\right) $. Suppose that
either $\#F\left( R^{\prime }\right) >1$ or [$\#F\left( R^{\prime }\right) =1$
and $F\left( R^{\prime }\right) \notin F\left( R\right) $]. Fix any $s_{i}\in
S_{j}$. We proceed according to whether $s_{i}\in MSS\left( \Gamma
,R^{\prime }\right) $ or not.
\medskip

\noindent {\bf Case 1}: $s_{i}\in MSS\left( \Gamma ,R^{\prime }\right) $
By the implementability of $F$, $h(s_i)\in F(R)\cap F(R')$. Since  by the assumption that $F\left( R^{\prime
}\right) \notin F\left( R\right) $ whenever $\#F\left( R^{\prime }\right) =1$, it must be  that $\#F\left( R^{\prime
}\right) >1$.  Since $\Gamma $ implements $F$ in rotation program, the set $%
MSS\left( \Gamma ,R^{\prime }\right) $ is partitioned in rotation programs $%
\left\{ \bar{S}_{1},...,\bar{S}_{m}\right\} $ such that $h\circ \bar{S}%
_{i}=F\left( R^{\prime }\right) $ for all $i=1,...,m$. Then, there exists a
unique $j$ such that $s_{i}\in \bar{S}_{j}$. Without loss of generality, let
$s_{i}=s_{1}\in \bar{S}_{j}$.

\medskip

\noindent {\bf Step 1:}.
Since $\bar{S}_{j}$ is a rotation program and since $\#F\left( R^{\prime
}\right) >1$, it follows that there exist $s_{2}\in \bar{S}_{j}\backslash
\left\{ s_{1}\right\} $ and a coalition $K_{1}$ such that $K_{1}\in \gamma
\left( s_{1},s_{2}\right) $ and $h\left( s_{2}\right) P_{K_{1}}^{\prime
}h\left( s_{1}\right) $.
Suppose that there exists $i_{1}\in K_{1}$ such that $h\left( s_{1}\right)
R_{i_{1}}h\left( s_{2}\right) $. Then, there exists $h\left( s_{2}\right)
\in Z$ such that $h\left( s_{2}\right) P_{i_{1}}^{\prime }h\left(
s_{1}\right) $ and $h\left( s_{1}\right) R_{i_{1}}h\left( s_{2}\right) $,
where $h\left( s_{1}\right) =h\left( s_{i}\right) =x\left( i,R\right) $.
Otherwise, suppose that $h\left( s_{2}\right) P_{K_{1}}h\left( s_{1}\right) $%
. Since $S_{j}$ is a rotation program, it follows that $s_{2}=s_{i+1}\in
S_{j}$ and $h\left( s_{i+1}\right) =x\left( i+1,R\right) $.

The above Step 1 can be applied to $s_{2}=s_{i+1}\in \bar{S}_{j}$ to derive
a state $s_{3}\in \bar{S}_{j}\backslash \left\{ s_{2}\right\} $ and a
coalition $K_{2}$ such that $K_{2}\in \gamma \left( s_{2},s_{3}\right) $ and
$h\left( s_{3}\right) P_{K_{2}}^{\prime }h\left( s_{2}\right) $ where $%
h\left( s_{2}\right) =x\left( i+1,R\right) $.
Suppose that $s_{3}=s_{1}$. Since $\bar{S}_{j}$ is a rotation program, it
follows that $\bar{S}_{j}=\left\{ s_{1},s_{2}\right\} $. Since $F\left(
R^{\prime }\right) \neq F\left( R\right) $, it follows that $s_{3}=s_{1}\neq
s_{i+2}\in S_{j}$. It follows that there exists $i_{2}\in K_{2}$ such that $%
h\left( s_{1}\right) P_{i_{2}}^{\prime }h\left( s_{2}\right) $ and $h\left(
s_{2}\right) R_{i_{2}}h\left( s_{1}\right) $. Thus, $zP_{i_{2}}^{\prime
}x\left( i+1,R\right) P_{i_{1}}^{\prime }x\left( i,R\right) $ and $x\left(
i+1,R\right) R_{i_{2}}z$ where $z=h\left( s_{1}\right) =x\left( i,R\right)
\in Z$.
Suppose that $s_{3}\neq s_1$. Then,  $s_{3}\in \bar{S}_{j}\backslash \left\{ s_{1},s_{2}\right\} $.
Suppose that there exists $i_{2}\in K_{2}$ such that $h\left( s_{2}\right)
R_{i_{2}}h\left( s_{3}\right) $. Thus, there exists $h\left( s_{3}\right)
=z\in Z$ such that
$
h\left( s_{3}\right) P_{i_{2}}^{\prime }h\left( s_{2}\right)
P_{i_{1}}^{\prime }h\left( s_{1}\right)
$
and
$
h\left( s_{2}\right) R_{i_{2}}h\left( s_{3}\right) \text{,}
$
where $h\left( s_{1}\right) =h\left( s_{i}\right) =x\left( i,R\right) $ and $%
h\left( s_{2}\right) =h\left( s_{i+1}\right) =x\left( i+1,R\right) $.
Otherwise, suppose that $h\left( s_{3}\right) P_{K_{2}}h\left( s_{2}\right) $%
. Since $S_{j}$ is a rotation program, it follows that $s_{3}=s_{i+2}\in
S_{j}$ and $h\left( s_{i+2}\right) =x\left( i+2,R\right) $. And, so on.

Since $\bar{S}_{j}\neq S_{j}$, after a finite number $1\leq h\leq m$ of
iterations, $s_{1},s_{2},...,s_{h+1}$ states and $i_{1},i_{2},..,i_{h}$
agents can be derived such that $s_{1},...,s_{h}\in \bar{S}_{j}\cap S_{j}$,
with $h\left( s_{\ell }\right) =h\left( s_{i+\ell -1}\right) =x\left( i+\ell
-1,R\right) $ for all $\ell =1,...,h$, $s_{h+1}\in \bar{S}_{j}$, $h\left(
s_{h+1}\right) =z\in Z$ and for all $\ell \in \left\{ 1,...,h\right\} $,%
$
h\left( s_{\ell +1}\right) P_{i_{\ell }}^{\prime }h\left( s_{\ell }\right)
\text{ and }h\left( s_{h}\right) R_{i_{h}}h\left( s_{h+1}\right) \text{.}
$
\medskip

\noindent  {\bf Case 2}: $s_{i}\notin MSS\left( \Gamma ,R^{\prime }\right)
$.
By iterated external stability of $MSS\left( \Gamma ,R^{\prime }\right) $,
there exists a finite myopic improvement path from $s_{i}$ to $t\in
MSS\left( \Gamma ,R^{\prime }\right) $; that is, there are coalitions $%
\{K_{1},...,K_{q-1}\}$ and states $\{s_{i}=t_{1},t_{2},...,t_{q}=t\}$ such that for all $%
p=1,...,q-1$,
$
K_{p}\in \gamma \left( t_{p},t_{p+1}\right)
$
and
$
h\left( t_{p+1}\right) P_{K_{p}}^{\prime }h\left( t_{p}\right) \text{.}
$
Since $\Gamma $ implements $F$ in rotation program, the set $MSS\left(
\Gamma ,R^{\prime }\right) $ is partitioned in rotation programs $\left\{
\bar{S}_{1},...,\bar{S}_{m}\right\} $ such that $h\circ \bar{S}_{i}=F\left(
R^{\prime }\right) $ for all $i=1,...,m$. Then, there exists a unique $j$
such that $t_{q}\in \bar{S}_{j}$.

\medskip

\noindent {\bf Step 1:} Suppose that $t_{2}\neq s_{i+1}$. Since $S_{j}$ is a rotation program and $%
s_{i}=t_{1}\in S_{j}$, it follows that there exists $i_{1}\in K_{1}$ such
that $h\left( t_{1}\right) R_{i_{1}}h\left( t_{2}\right) $ where $h\left(
t_{1}\right) =h\left( s_{i}\right) =x\left( i,R\right) $. Therefore, $%
h\left( t_{2}\right) P_{i_{1}}^{\prime }h\left( t_{1}\right) $ and $h\left(
t_{1}\right) R_{i_{1}}h\left( t_{2}\right) $, as we sought.
Otherwise, suppose that $t_{2}=s_{i+1}\in S_{j}$. If there exists $i_{1}\in
K_{1}$ such that $h\left( t_{1}\right) R_{i_{1}}h\left( t_{2}\right) $, then
again $h\left( t_{2}\right) P_{i_{1}}^{\prime }h\left( t_{1}\right) $ and $%
h\left( t_{1}\right) R_{i_{1}}h\left( t_{2}\right) $. Otherwise, suppose
that $t_{2}=s_{i+1}\in S_{j}$, $h\left( t_{2}\right) =x\left( i+1,R\right) $
and $h\left( t_{2}\right) P_{K_{1}}h\left( t_{1}\right) $.

The reasoning used in the above Step 1 can be applied to $t_{3}$ to conclude
that either there exists $i_{2}\in K_{2}$ such that $h\left( t_{2}\right)
R_{i_{2}}h\left( t_{3}\right) $ for some $i_{2}\in K_{2}$ or $h\left(
t_{3}\right) P_{K_{2}}h\left( t_{2}\right) $ and $t_{3}=s_{i+2}\in S_{j}$.

In the former case, we have that
$
h\left( t_{3}\right) P_{i_{2}}^{\prime }h\left( t_{2}\right)
P_{i_{1}}^{\prime }h\left( t_{1}\right) \text{ and }h\left( t_{2}\right)
R_{i_{2}}h\left( t_{3}\right) \text{,}
$
where $h\left( t_{1}\right) =x\left( i,R\right) $ and $h\left( t_{2}\right)
=x\left( i+1,R\right) $. In the latter case, we have that $h\left(
t_{3}\right) =x\left( i+2,R\right) $ and $h\left( t_{3}\right)
P_{K_{2}}h\left( t_{2}\right) $.

Since the myopic improvement path from $s_{i}$ to $t\in MSS\left( \Gamma
,R^{\prime }\right) $ is finite, after a finite number $1\leq r\leq q-1$ of
iterations, we have that $h\left( t_{p+1}\right) P_{i_{p}}^{\prime }h\left(
t_{p}\right) $ for all $p=1,...,r$, and either [$h\left( t_{r}\right)
R_{i_{r}}h\left( t_{r+1}\right) $ for some $i_{r}\in K_{r}$] or [$r=q-1$, $%
h\left( t_{p+1}\right) P_{K_{p}}h\left( t_{p}\right) $ and $%
t_{p}=s_{i+p-1}\in S_{j}$ for all $p=1,...,r$, and $t_{q}\in S_{j}\cap \bar{S%
}_{j}$]. In the former case, we have that for all $p=1,...,r$,
$
h\left( t_{p+1}\right) P_{i_{p}}^{\prime }h\left( t_{p}\right) \ \text{
and}\ h\left( t_{r}\right) R_{i_{r}}h\left( t_{r+1}\right) \text{,}
$
where $h\left( t_{p}\right) =h\left( s_{i+p-1}\right) =x\left( i+p-1\right) $
for all $p=1,...,r$. In the latter case, since $t_{q}\in \bar{S}_{j}$, it
follows that $t_{q}\in MSS\left( \Gamma ,R^{\prime }\right) $. Case 1 above
can be applied to the outcome $h\left( t_{q}\right) =h\left(
s_{i+q-1}\right) =x\left( i+q-1\right) \in F\left( R\right) $ to complete
the proof.
\medskip

\noindent \textbf{Proof of  \Cref{Th4}}. The
implementing rights structure is a variant of the rights structure
constructed in the proof of \Cref{Th1}. What changes is only the definition
of Rule 1. The state space is $S=Gr\left( F\right) \cup Z$. The outcome
function is $h\left( x,R\right) =x$ for all $\left( x,R\right) \in Gr\left(
F\right) $ and $h\left( x\right) =x$ for all $x\in Z$. The code of rights $%
\gamma $ is defined as follows. For all $i\in N$, all $R\in \mathcal{R}$ and
all $s,t\in S$:

\medskip

\noindent \textbf{RULE\ 1:} If $s=\left( x\left( k,R\right) ,R\right) $ and $%
t=\left( x\left( k+1,R\right) ,R\right) $ for some $1\leq k\leq m$, then
$
\left\{ i\right\}\in \gamma \left( \left( x\left( k,R\right) ,R\right) ,\left( x\left(
k+1,R\right) ,R\right) \right) \text{,}
$ where the outcomes $x\left( k,R\right) $ are those specified by properties 1 and 2.

\medskip

\noindent \textbf{RULE\ 2:} If $s=\left( z,R\right) $, $t=x$ and $x\in
L_{i}\left( z,R\right) $, then $\left\{ i\right\} \in \gamma \left( \left(
z,R\right) ,x\right) $.

\medskip

\noindent \textbf{RULE\ 3:} If $s=x$ and $t=\left( z,R\right) $, then $%
\left\{ i\right\} \in \gamma \left( x,\left( z,R\right) \right) $.

\medskip

\noindent \textbf{RULE\ 4:} If $s=z$ and $t=x$, then $\left\{ i\right\} \in
\gamma \left( s,t\right) $.

\medskip

\noindent \textbf{RULE\ 5:} Otherwise, $\gamma \left( s,t\right) =\emptyset $%
.

\medskip

Rule 1 allows agent $i$ to be effective only between two consecutive 
socially optimal outcomes at $R$, that is, between $\left( x\left( k,R\right) ,R\right) $ and $%
\left( x\left( k+1,R\right) ,R\right) $ for all  $1\leq k\leq m$. 
Fix any $R$. Let us show that $\Gamma $ implements $F$ in rotation programs. We first show that $F\left( R\right) =h\circ MSS(\Gamma ,R) $ and then we show that $\Gamma $ partitions $MSS\left( \Gamma
,R\right) $ in rotation programs such that for each rotation program $S$, it holds that  $F\left( R\right) =h\circ S $.
To show that $F\left( R\right) =h\circ MSS\left( \Gamma ,R\right) $ and that
$MSS\left( \Gamma ,R\right) =M\left( R\right) \cup U\left( R\right) \cup
Q\left( R\right) $, we need to show that Lemmata 1-7 still hold under the
new rights structure $\Gamma $. It can be checked that the only proofs that
need to be amended are the proofs of \Cref{lem2} and \Cref{lem3}.
As far as the proof of \Cref{lem3} is concerned, the arguments provided to prove
Step 2 of  \Cref{lem3}  no longer hold. However, the statement of this step is still true
under the new $\Gamma $. To show this, take any $s=(x(i,R),R)\in M(R)\cap V$%
, which exists by Step 1 of the proof of \Cref{lem3}. We show that $%
M(R)\subseteq V$. Assume, to the contrary, there exists $s^{\prime
}=(x(i^{\prime },R),R)\in M(R)$ such that $s^{\prime }\notin V$.
To complete the proof of \Cref{lem3}, let us first show that $M(R)$ is a rotation program. Since $F$ is efficient and since $\mathcal{R}$ satisfies the restriction in (\ref%
{pref-dom}), it follows that for all $1\leq k\leq m$ and all $\left( x\left(
k,R\right) ,R\right) ,\left( x\left( k+1,R\right) ,R\right) \in M\left(
R\right) $, there exists $j\in N$ such that $x\left( k+1,R\right)
P_{j}x\left( k,R\right) $. By definition of Rule 1, it follows that for each
$1\leq k\leq m$, there exists $j\in N$ such that $\left\{ j\right\} \in
\gamma \left( \left( x\left( k,R\right) ,R\right) ,\left( x\left(
k+1,R\right) ,R\right) \right) $ and $x\left( k+1,R\right) P_{j}x\left(
k,R\right) $. Moreover, by definition of $\gamma $, it follows that $M\left(
R\right) $ is a rotation program because for each $\left( x\left( k,R\right)
,R\right) $, there do not exist any $K\in \mathcal{N}_{0}$ and any $s\in S$,
with $s\neq \left( x\left( k,R\right) ,R\right) $ and $s\neq \left( x\left(
k+1,R\right) ,R\right) $, such that $K\in \gamma \left( \left( x\left(
k,R\right) ,R\right) ,s\right) $ and $h\left( s\right) P_{K}x\left(
k,R\right) $.
Let us now complete the proof of \Cref{lem3}. Since for each $1\leq k\leq m$ there exists $j\in N$ such that $\left\{
j\right\} \in \gamma (\left( x\left( k,R\right) ,R\right) ,$ $\left( x\left(
k+1,R\right) ,R\right) )$ and $x\left( k+1,R\right) P_{j}x\left( k,R\right) $%
, it follows that there exist $s_{0},s_{1},...,$ $s_{p-1,}s_{p}$, with $%
s_{0}=s$ and $s_{p}=s^{\prime }$, and $i_{0},...,i_{p-1}$ such that $%
i_{h}\in \gamma \left( s_{h},s_{h+1}\right) $ and $h\left( s_{h+1}\right)
P_{i_{h}}h\left( s_{h}\right) $ for all $h=0,...,p-1$, where $s_{h}\in
M\left( R\right) $ for all $h=0,1,...,p$. Since $s_{0}\in M(R)\cap V$ and $%
s_{p}\in M\left( R\right) \backslash V$, there exists the smallest index $%
h^{\ast }\in \left\{ 0,...,p-1\right\} $ such that $s_{h^{\ast }}\in
M(R)\cap V$ and $s_{h^{\ast }+1}\in M\left( R\right) \backslash V$. Since $%
i_{h^{\ast }}\in \gamma \left( s_{h^{\ast }},s_{h^{\ast }+1}\right) $ and $%
h\left( s_{h^{\ast }+1}\right) P_{i_{h^{\ast }}}h\left( s_{h^{\ast }}\right)
$, this contradicts our initial supposition that $V$ satisfies the property
of deterrence of external deviations. Thus, we have that $M\left( R\right)
\subseteq V$, and so \Cref{lem3} holds as well.

As far as the proof of \Cref{lem2} is concerned, it needs to be amended as
follows. Fix any $R^{\prime }\in \mathcal{R}$. The proof of \Cref{lem2} holds if
$\#F\left( R\right) \neq 1$ or if $\#F\left( R\right) =1$ and $F\left(
R\right) \notin F\left( R^{\prime }\right) $. The reason is that in these
cases {\it rotation monotonicity} implies {\it indirect monotonicity}. To complete the proof of
\Cref{lem2}, let us suppose that $\#F\left( R\right) =1$ and $F\left( R\right)
\in F\left( R^{\prime }\right) $.
Suppose that $F\left( R\right) =\left\{ a\right\} \neq F\left( R^{\prime
}\right) =\left\{ z\left( 1,R^{\prime }\right) ,...,z\left( m,R^{\prime
}\right) \right\} $. Without loss of generality, let $a=z\left( 1,R^{\prime
}\right) $.
Suppose that {\it Property M} implies that for each $z\left( i,R^{\prime }\right)
\in F\left( R^{\prime }\right) \backslash \left\{ z\left( 1,R^{\prime
}\right) \right\} $, there exist $x\in Z$ and $i_{1},...,i_{h}$, with $1\leq
h\leq m$, such that:%
\[
z\left( i+\ell +1,R^{\prime }\right) P_{\ell +1}z\left( i+\ell ,R^{\prime
}\right) \text{ for all }\ell \in \left\{ 0,...,h-1\right\} \ \ \text{and}
\]%
\[
z\left( i+h,R^{\prime }\right) P_{h}x\text{ and }xR_{h}^{\prime }z\left(
i+h,R^{\prime }\right) \text{.}
\]

By definition of $\gamma $, we have that for each $z\left( i,R^{\prime
}\right) \in F\left( R^{\prime }\right) \backslash \left\{ z\left(
1,R^{\prime }\right) \right\} $, there exists a finite myopic improvement
path from $\left( z\left( i,R^{\prime }\right) ,R^{\prime }\right) $ to $x$.
Suppose that $U\left( R\right) \neq \emptyset $. Since $F$ is efficient and
since, moreover, $\mathcal{R}$ satisfies the restriction in (\ref%
{pref-dom}), it follows that $U\left( R\right) =\left\{
z\left( 1,R^{\prime }\right) \right\} $. Since by Rule 2 there exists a
finite myopic improvement path from $x$ to $z\left( 1,R^{\prime }\right) $,
it follows that there exists a finite myopic improvement path from $z\left(
i,R^{\prime }\right) \in F\left( R^{\prime }\right) \backslash \left\{
z\left( 1,R^{\prime }\right) \right\} $ to $M\left( R\right) \cup U\left(
R\right) $. Suppose that $U\left( R\right) =\emptyset $. Since \Cref{lem1}
implies that there exists a finite myopic improvement path from $x$ to $%
M\left( R\right) \cup U\left( R\right) $, we conclude that there exists a
finite myopic improvement path from $z\left( i,R^{\prime }\right) \in
F\left( R^{\prime }\right) \backslash \left\{ z\left( 1,R^{\prime }\right)
\right\} $ to $M\left( R\right) \cup U\left( R\right) $. It follows from the
definition of $Q\left( R,R^{\prime }\right) \subseteq M\left( R^{\prime
}\right) $ that $Q\left( R,R^{\prime }\right) =\emptyset $ if there exists a
finite myopic improvement path from $\left( z\left( 1,R^{\prime }\right)
,R^{\prime }\right) $ to $M\left( R\right) \cup U\left( R\right) $,
otherwise, $Q\left( R,R^{\prime }\right) =\left\{ \left( z\left( 1,R^{\prime
}\right) ,R^{\prime }\right) \right\} $. In either case, we have that $%
h\circ Q\left( R,R^{\prime }\right) \subseteq F\left( R\right) $ and that $%
Q\left( R,R^{\prime }\right) $ satisfies the property of deterrence of
external deviations. Note that $Q\left( R,R^{\prime }\right) =\left\{ \left( z\left( 1,R^{\prime
}\right) ,R^{\prime }\right) \right\} $ satisfies this property for the
following two reasons: 1) Since every agent $i$ is effective in move the
state from $\left( z\left( 1,R^{\prime }\right) ,R^{\prime }\right) $ to $%
\left( z\left( 2,R^{\prime }\right) ,R^{\prime }\right) $, it cannot be that
$z\left( 2,R^{\prime }\right) P_{i}z\left( 1,R^{\prime }\right) $ for some $i
$, otherwise, since we have already shown that there exists a finite myopic
improvement path from $\left( z\left( 1,R^{\prime }\right) ,R^{\prime
}\right) $ to $M\left( R\right) \cup U\left( R\right) $, it follows that $%
Q\left( R,R^{\prime }\right) =\emptyset $, which is a contradiction; and 2)\
it cannot be that $xP_{i}z\left( 1,R^{\prime }\right) $ for some $i$ and
some $x\in L_{i}\left( z\left( 1,R^{\prime }\right) ,R^{\prime }\right) $,
otherwise, since Rule 2 implies that $\left\{ i\right\} \in \gamma \left(
\left( z\left( 1,R^{\prime }\right) ,R^{\prime }\right) ,x\right) $ and $%
xP_{i}z\left( 1,R^{\prime }\right) $ and since, moreover, \Cref{lem1} implies
that there exists a finite myopic improvement path from $x$ to $M\left(
R\right) \cup U\left( R\right) $, since we have already shown that there
exists a finite myopic improvement path from $\left( z\left( 1,R^{\prime
}\right) ,R^{\prime }\right) $ to $M\left( R\right) \cup U\left( R\right) $,
it follows that $Q\left( R,R^{\prime }\right) =\emptyset $, which is a
contradiction.
Suppose that the above arguments do not hold for some $z\left( i,R^{\prime
}\right) \in F\left( R^{\prime }\right) \backslash \left\{ z\left(
1,R^{\prime }\right) \right\} $. Clearly, for each $z\left( i,R^{\prime
}\right) \in F\left( R^{\prime }\right) \backslash \left\{ z\left(
1,R^{\prime }\right) \right\} $ such that the above arguments hold, we have
that there exists a finite myopic improvement path from $z\left( i,R^{\prime
}\right) \in F\left( R^{\prime }\right) \backslash \left\{ z\left(
1,R^{\prime }\right) \right\} $ to $M\left( R\right) \cup U\left( R\right) $%
. {\it Property M} implies that $L_{i}\left( z\left( 1,R^{\prime }\right)
,R^{\prime }\right) \cup \left\{ z\left( 2,R^{\prime }\right) \right\}
\subseteq L_{i}\left( z\left( 1,R^{\prime }\right) ,R\right) $ for all $i\in
N$. For each $z\left( i,R^{\prime }\right) \in F\left( R^{\prime }\right)
\backslash \left\{ z\left( 1,R^{\prime }\right) \right\} $ for which the
above arguments do not hold, {\it Property M} implies that there exists a sequence
of agents $i_{1},...,i_{\ell }$ such that%
\begin{equation}
z\left( 1,R^{\prime }\right) P_{i_{\ell }}z\left( m,R^{\prime }\right)
P_{i_{\ell -1}}\cdot \cdot \cdot P_{i_{2}}z\left( i+1,R^{\prime }\right)
P_{i_{1}}z\left( i,R^{\prime }\right)   \label{a}
\end{equation}

\noindent Since every agent $i$ can be effective in moving the state from $\left(
z\left( 1,R^{\prime }\right) ,R^{\prime }\right) $ to $\left( z\left(
2,R^{\prime }\right) ,R^{\prime }\right) $, it follows that no agent has an
incentive to do so because $z\left( 2,R^{\prime }\right) \in L_{i}\left(
z\left( 1,R^{\prime }\right) ,R\right) $ for all $i\in N$. Since, by Rule 1,
each agent $i\in \left\{ i_{1},...,i_{\ell }\right\} $ is effective in
moving between two consecutive states in $M\left( R^{\prime }\right) $, it
follows from (\ref{a}) that there exists a finite
myopic improvement path from $\left( z\left( i,R^{\prime }\right) ,R^{\prime
}\right) $ to $\left( z\left( 1,R^{\prime }\right) ,R^{\prime }\right) $. We
conclude that for each $z\left( i,R^{\prime }\right) \in F\left( R\right)
\backslash \left\{ z\left( 1,R^{\prime }\right) \right\} $, there exists a
finite myopic improvement path from $\left( z\left( i,R^{\prime }\right)
,R^{\prime }\right) $ to either $M\left( R\right) \cup U\left( R\right) $ or
to $\left\{ \left( z\left( 1,R^{\prime }\right) ,R^{\prime }\right) \right\}
$.
It follows that $Q\left( R,R^{\prime }\right) \subseteq \left\{ \left(
z\left( 1,R^{\prime }\right) ,R^{\prime }\right) \right\} $. Again, $Q\left(
R,R^{\prime }\right) =\emptyset $ if there exists a finite myopic
improvement path from $\left( z\left( 1,R^{\prime }\right) ,R^{\prime
}\right) $ to $M\left( R\right) \cup U\left( R\right) $, otherwise, $Q\left(
R,R^{\prime }\right) =\left\{ \left( z\left( 1,R^{\prime }\right) ,R^{\prime
}\right) \right\} $. In either case, we have that $h\circ Q\left(
R,R^{\prime }\right) \subseteq F\left( R\right) $ and that $Q\left(
R,R^{\prime }\right) $ satisfies the property of deterrence of external
deviations.
Since the choice of $R^{\prime }\in \mathcal{R}$ is arbitrary, it follows
that \Cref{lem2} holds.
 Since Properties 1-2 implies that Lemmata 1-7 hold, it follows that $F\left(
R\right) =h\circ MSS\left( \Gamma ,R\right) $ and that $MSS\left( \Gamma
,R\right) =M\left( R\right) \cup U\left( R\right) \cup Q\left( R\right) $.

To show that $\Gamma $ partitions $MSS\left( \Gamma ,R\right) $ in rotation
programs, we proceed according to whether $\#F\left( R\right) =1$ or not. We have already shown above that $M(R)$ is a rotation program.
\medskip

\noindent {\bf Case 1}: $\#F\left( R\right) \neq 1$.
The set $U\left( R\right) =\emptyset $. To see it, suppose that there exists
$x\in U\left( R\right) $. Since $F$ is efficient and since, moreover, $%
\mathcal{R}$ satisfies the restriction in (\ref{pref-dom}), it follows that $%
F\left( R\right) =\left\{ x\right\} $, which is a contradiction. Thus, $%
MSS\left( \Gamma ,R\right) =M\left( R\right) \cup Q\left( R\right) $. We
have already shown above that $M\left( R\right) $ is a rotation program.
Moreover, by its definition, it follows that $F\left( R\right) =h\circ
M\left( R\right) $.

Fix any $R^{\prime }\in \mathcal{R}$ such that $F\left( R^{\prime }\right)
\neq F\left( R\right) $. We show that $Q\left( R,R^{\prime }\right)
=\emptyset $. Fix any $z\left( i,R^{\prime }\right) \in F\left( R^{\prime
}\right) $. {\it Rotation monotonicity} implies that there exist $x\in Z$ and a sequence of
agents $i_{1},...,i_{h}$, with $1\leq h\leq m$, such that:%
\[
z\left( i+\ell +1,R^{\prime }\right) P_{i_{\ell +1}}z\left( i+\ell
,R^{\prime }\right) \text{ for all }\ell \in \left\{ 0,...,h-1\right\}\ \  \text{and}
\]%
\[
z\left( i+h,R^{\prime }\right) R_{i_{h}}^{\prime }x\text{ and }%
xP_{i_{h}}z\left( i+h,R^{\prime }\right) \text{.}
\]%
Since, by Rule 1, for each $\ell \in \left\{ 0,...,h-1\right\} $, $\left\{
i_{\ell +1}\right\} \in \gamma (z\left( i+\ell ,R^{\prime }\right) ,$ $%
z(i+\ell +1,$ $R^{\prime }))$ and since, moreover, by Rule 2, $\left\{
i_{h}\right\} \in \gamma \left( z\left( i+h,R^{\prime }\right) ,x\right) $,
it follows that there exists a finite myopic improvement path from $\left(
z\left( i,R^{\prime }\right) ,R^{\prime }\right) $ to $x$. Since $U\left(
R\right) =\emptyset $, \Cref{lem1} implies that there exists a finite myopic
improvement path from $x$ to $M\left( R\right) $. Therefore, we have
established that there exists a finite myopic improvement path from $\left(
z\left( i,R^{\prime }\right) ,R^{\prime }\right) $ to $M\left( R\right) $,
and so $\left( z\left( i,R^{\prime }\right) ,R^{\prime }\right) \notin
Q\left( R,R^{\prime }\right) $. Since the choice of $z\left( i,R^{\prime
}\right) \in F\left( R^{\prime }\right) $ is arbitrary, we have that $%
Q\left( R,R^{\prime }\right) =\emptyset $.

Fix any $R^{\prime }\in \mathcal{R}$ such that $F\left( R^{\prime }\right)
=F\left( R\right) $. Nothing has to be proved if $Q\left( R,R^{\prime }\right) =\emptyset $.
Suppose that $Q\left( R,R^{\prime }\right) \neq \emptyset $. We show that $%
Q\left( R,R^{\prime }\right) =M\left( R^{\prime }\right) $ and that $Q\left( R,R^{\prime }\right)$ is a rotation program. Since $F$ is
efficient and since $\mathcal{R}$ satisfies the restriction in (\ref%
{pref-dom}), it follows that for all $\left( x\left( k,R^{\prime }\right)
,R^{\prime }\right) ,(x\left( k+1,R^{\prime }\right) ,$ $R^{\prime })\in
M\left( R^{\prime }\right) $, there exists $j\in N$ such that $x\left(
k+1,R^{\prime }\right) P_{j}x\left( k,R^{\prime }\right) $. By definition of
Rule 1, it follows that for each $1\leq k\leq m$, there exists $j\in N$ such
that $\left\{ j\right\} \in \gamma \left( \left( x\left( k,R^{\prime
}\right) ,R^{\prime }\right) ,\left( x\left( k+1,R^{\prime }\right)
,R^{\prime }\right) \right) $ and $x\left( k+1,R^{\prime }\right)
P_{j}x\left( k,R^{\prime }\right) $. 
If there exists a finite myopic improvement path from some $\left( x\left(
i,R^{\prime }\right) ,R^{\prime }\right) \in M\left( R^{\prime }\right)
\backslash Q\left( R,R^{\prime }\right) $ to $M\left( R\right) \cup U\left(
R\right) $, it follows that for each state in $M\left( R^{\prime }\right) $
there exists a finite myopic improvement path to $M\left( R\right) \cup
U\left( R\right) $. This implies that $Q\left( R,R^{\prime }\right)
=\emptyset $, which is a contradiction. Thus, $Q\left(
R,R^{\prime }\right) =M\left( R^{\prime }\right) $.
Since \Cref{lem2} implies that $Q\left( R,R^{\prime }\right) $ satisfies the
property of deterrence of external deviations, it follows that $Q\left( R,R^{\prime }\right) $ is a rotation program.
Since the choice of $R^{\prime }\in \mathcal{R}$, with $F\left( R^{\prime
}\right) =F\left( R\right) $, is arbitrary, it follows that $MSS\left(
\Gamma ,R\right) $ is the union of partitioned rotation programs because for
all $R^{\prime },R^{\prime \prime }\in \mathcal{R}$ such that $F\left(
R^{\prime }\right) =F\left( R^{\prime \prime }\right) =F\left( R\right) $,
it holds that $h\circ M\left( R^{\prime }\right) =h\circ M\left( R^{\prime
\prime }\right) $ and $M\left( R^{\prime }\right) \cap M\left( R^{\prime
\prime }\right) =\emptyset $. Thus, $F$ is rotationally programmatically
implementable.
\medskip

\noindent  {\bf Case 2}: $\#F\left( R\right) =1$.
Recall that $MSS\left( \Gamma ,R\right) =M\left( R\right) \cup U\left(
R\right) \cup Q\left( R\right) $. Let $F\left( R\right) =\left\{ z\left( 1,R\right)\right\} $%
. Note that $M\left( R\right) =\left( z\left( 1,R\right) ,R\right) $. Also,
note that if $U\left( R\right) \neq \emptyset $, it follows from the
efficiency of $F$ and the restriction of $\mathcal{R}$ in (\ref{pref-dom})
that $U\left( R\right) =\left\{ z\left( 1,R\right)\right\} $. Note that $M\left( R\right) $
and $U\left( R\right) $ are rotation programs such that $M\left( R\right)
\cap U\left( R\right) =\emptyset $. To proof is complete if we show that for all $R^{\prime }\in \mathcal{R}$,
either $Q\left( R,R^{\prime }\right) =\emptyset $ or $Q\left( R,R^{\prime
}\right) =\left\{ \left( z\left( 1,R\right),R^{\prime }\right) \right\} $. To this end, fix any $%
R^{\prime }\in \mathcal{R}$.
Suppose that $F\left( R\right) =\left\{ z\left( 1,R\right)\right\} \neq F\left( R^{\prime
}\right) $. Let us proceed according whether $F\left( R\right) \in F\left(
R^{\prime }\right) $ or not.
Suppose that $F\left( R\right) \notin F\left( R^{\prime }\right) $. Fix any $%
z\left( i,R^{\prime }\right) \in F\left( R^{\prime }\right) $. By the same
arguments provided in Case 1 above, it follows that there exists a finite
myopic improvement path from $\left( z\left( i,R^{\prime }\right) ,R^{\prime
}\right) $ to $x$. If $U\left( R\right) \neq \emptyset $, then there exists
a finite myopic improvement path from $\left( z\left( i,R^{\prime }\right)
,R^{\prime }\right) $ to $z\left( 1,R\right)\in U\left( R\right) $. Otherwise, if $U\left(
R\right) =\emptyset $, \Cref{lem1} implies that there exists a finite myopic
improvement path from $x$ to $M\left( R\right) $. Therefore, there exists a
finite myopic improvement path from $\left( z\left( i,R^{\prime }\right)
,R^{\prime }\right) $ to $M\left( R\right) \cup U\left( R\right) $, and so $%
\left( z\left( i,R^{\prime }\right) ,R^{\prime }\right) \notin Q\left(
R,R^{\prime }\right) $. Since the choice of $z\left( i,R^{\prime }\right)
\in F\left( R^{\prime }\right) $ is arbitrary, we have that $Q\left(
R,R^{\prime }\right) =\emptyset $.
Suppose that $F\left( R\right) \in F\left( R^{\prime }\right) =\left\{
z\left( 1,R^{\prime }\right) ,...,z\left( m,R^{\prime }\right) \right\} $.
Without loss of generality, suppose that $z\left( 1,R\right)=z\left( 1,R^{\prime }\right) $. By arguing as we have done above in the completion of the proof of \Cref{lem2},
we have that either $Q\left( R,R^{\prime }\right) =\emptyset $ or $Q\left(
R,R^{\prime }\right) =\left\{ \left( z\left( 1,R^{\prime }\right) ,R^{\prime
}\right) \right\} $, as we sought.\hfill $\blacksquare$
\medskip

\noindent \textbf{Proof of \Cref{Th5}.} In light of
\Cref{Th3}, it suffices to show that $F$ satisfies properties 1 and 2. Since $%
\#F\left( R\right) >1$ for all $R\in \mathcal{\bar{R}}$, it follows that
{\it Property M} is vacuously satisfied. Therefore, let us show that $F$ satisfies
{\it rotation monotonicity} as well. To this end, we need to introduce additional notation.

For all $R\in \mathcal{\bar{R}}$ and all $i\in N$, let $N_{i}\left( R\right)
$ denote the set of Pareto efficient allocations at $R$ that assign $%
j_{1}^{\ast }$ to agent $i$, with $n_{i}\left( R\right) $ representing the
number of elements in $N_{i}\left( R\right) $. Since $J$ is a finite set, it
follows that $N_{i}\left( R\right) $ is a finite set. For all $R\in \mathcal{%
\bar{R}}$ and all $i\in N$, let $\tau _{2}\left( i,R\right) $ denote the
second top-ranked job of agent $i$ at $R_{i}$. For all $x\in \bar{J}$ and
all $R\in \mathcal{\bar{R}}$, let $\bar{x}\left( R\right) $ be a permutation
of $x$ such that (i) the agent who obtains $j_{1}^{\ast }$ at $x$, let us
say agent $i$, obtains his second top-ranked job $\tau _{2}\left( i,R\right)
$ at $\bar{x}\left( R\right) $; (ii) the agent who obtains agent $i$'s
second top-ranked job at $x$ obtains $j_{1}^{\ast }$ at $\bar{x}\left(
R\right) $; whereas (iii) all other agents obtain the same job both at $x$
and at $\bar{x}\left( R\right) $. Formally, $\bar{x}_{i}\left( R\right)
=\tau _{2}\left( i,R\right) $ if $x_{i}=j_{1}^{\ast }$, $\bar{x}_{j}\left(
R\right) =j_{1}^{\ast }$ if $x_{j}=\tau _{2}\left( i,R\right) $, and $x_{h}=%
\bar{x}_{h}\left( R\right) $ for all $h\in N\backslash \left\{ i,j\right\} $.

The proof that $F$ satisfies {\it rotation monotonicity} relies on the following lemmata.

\begin{lemma}
\label{Lem1J*}For all $R\in \mathcal{\bar{R}}$ and all $i\in N$,$
\sum\limits_{j\in N\backslash \left\{ i\right\} }n_{j}\left( R\right) \geq
n_{i}\left( R\right) \text{.}  \label{AA}
$
\end{lemma}

\noindent {\bf Proof of \Cref{Lem1J*}:}
The statement follows if we show that for all $R\in \mathcal{\bar{R}}$ and
all $i\in N$, there exists an injective function $g_{i}^{R}$ from $%
N_{i}\left( R\right) $ to $\bigcup _{j\in N\backslash \left\{ i\right\}
}N_{j}\left( R\right) $, that is, if we show that for for all $R\in \mathcal{%
\bar{R}}$ and all $i\in N$, every two distinct elements of $N_{i}\left(
R\right) $ have distinct images in $\bigcup _{j\in N\backslash \left\{
i\right\} }N_{j}\left( R\right) $ under $g_{i}^{R}$. Let us define $%
g_{i}^{R}:N_{i}\left( R\right) \longrightarrow \bigcup _{j\in N\backslash
\left\{ i\right\} }N_{j}\left( R\right) $ by $g_{i}^{R}\left( x\right) =\bar{%
x}\left( R\right) $. Take any two distinct $x,y\in N_{i}\left( R\right) $.
Then, $g_{i}^{R}\left( x\right) =\bar{x}\left( R\right) $ and $%
g_{i}^{R}\left( y\right) =\bar{y}\left( R\right) $. Suppose that $%
x_{j}=y_{j}=\tau _{2}\left( i,R\right) $ for some $j\in N\backslash \left\{
i\right\} $. Since $x\neq y$, it follows that $x_{h}\neq y_{h}$ for some $%
h\in N\backslash \left\{ i,j\right\} $. It follows that $\bar{x}\left(
R\right) \neq \bar{y}\left( R\right) $. Suppose that $x_{j}=\tau _{2}\left(
i,R\right) $ and $y_{h}=\tau _{2}\left( i,R\right) $ for some $h,j\in
N\backslash \left\{ i\right\} $ such that $h\neq j$. It follows that $\bar{x}%
\left( R\right) \neq \bar{y}\left( R\right) $. Thus, $g_{i}^{R}$ is an
injective function. $\hfill \square$

\begin{lemma}
\label{Lem2J*}For all $R\in \mathcal{\bar{R}}$, elements of $F\left(
R\right) $ can be ordered as $x\left( 1,R\right) ,...,x\left( m,R\right) $,
with $m=\sum_{i\in N}n_{i}\left( R\right) >1$, such that for all $k=1,...,m$
(mod $m$), if $x_{i}\left( k,R\right) =j_{1}^{\ast }$ for some $i\in N$,
then $x_{i}\left( k+1,R\right) \neq j_{1}^{\ast }$.
\end{lemma}

\noindent{\bf Proof of \Cref{Lem2J*}:}
Fix any $R\in \mathcal{\bar{R}}$. Without loss of generality, let us assume
that%
$
n_{1}\left( R\right) \geq n_{2}\left( R\right) \geq ...\geq n_{n-1}\left(
R\right) \geq n_{n}\left( R\right) \text{.}
$
Let us apply the following procedure to arrange allocations of $F\left(
R\right) $ in a way that the statement holds:
\medskip

\noindent {\bf Step 0:} If $n_{1}\left( R\right) -n_{2}\left( R\right) =0$, then go
to Step 1. If $n_{1}\left( R\right) -n_{2}\left( R\right) =k_{0}>0$, then
take any $A\subseteq N_{1}\left( R\right) $ such that $\#A=k_{0}$. By  \Cref{Lem1J*}, there exists $3\leq h\leq n$ such that $\sum_{i=h}^{n}n_{i}\left(
R\right) \geq k_{0}$ and $\sum_{i=h+1}^{n}n_{i}\left( R\right) <k_{0}$. Then, select
any $B\subseteq N_{h}\left( R\right) $ such that $\sum_{i=h+1}^{n}n_{i}%
\left( R\right) +\#B=k_{0}$. List elements of the set $A$ and elements of the
set $B\cup \left( \cup _{i=h+1}^{n}N_{i}\left( R\right) \right) $ in a way
that no element of $A$ stands next to another element of set $A$. Start the
list with an element of $A\subseteq N_{1}\left( R\right) $. By construction,
no two consecutive allocations of the list allocate $j_{1}^{\ast }$ to the
same agent.
\medskip

\noindent {\bf Step 1:} Then, $n_{1}\left( R\right) -k_{0}-n_{2}\left( R\right) =0$,
with $k_{0}-0$ if $n_{1}\left( R\right) =n_{2}\left( R\right) $, and that
$
n_{1}\left( R\right) -k_{0}=n_{2}\left( R\right) \geq ...\geq n_{h}\left(
R\right) -\#B\text{,}
$
where $B=\varnothing $ and $h=n$ if $n_{1}\left( R\right) =n_{2}\left(
R\right) $. Let $n_{h}\left( R\right) -\#B=k_{1}$. Construct a sequence $%
\left\{ x_{i}\right\} _{i=1}^{h}$ of elements in $\bigcup _{i=1}^{h}N_{i}\left(
R\right) \backslash \left( A\cup B\right) $ (of length equal to $h$) such
that $x_{i}\in N_{i}\left( R\right) $ for all $i=1,...,h$. Thus, the
sequence is constructed in a way that that no element of $N_{i}\left(
R\right) $ stands next to another element of $N_{i}\left( R\right) $, and
the last element of the sequence belongs to $N_{h}\left( R\right) $. Since
there are $k_{1}$ sequences of this type, list these sequences one after the
other. By construction, no two consecutive allocations of this arrangement
allocate $j_{1}^{\ast }$ to the same agent. Join this linear arrangement to
the right end of the arrangement of Step 0. If $n_{h}\left( R\right)
-\#B=n_{1}\left( R\right) -k_{0}$, then the derived linear arrangement can
be transformed into a circular arrangement by joining its ends. Otherwise,
move to Step 2. For each $i=1,...,h-1$, let $A_{1i}$ denote the set of
elements of $N_{i}\left( R\right) $ used to construct the sequences. Thus,
for each $i=1,...,h-1$, $\#A_{1i}=k_{1}$ and $N_{i}\left( R\right)
\backslash A_{1i}$ is the set of allocations that still needs to be arranged.
\medskip 

\noindent {\bf Step 2:} Then, 
$
n_{1}\left( R\right) -k_{0}-k_{1}=n_{2}\left( R\right) -k_{1}\geq ...\geq
n_{h-1}\left( R\right) -k_{1}\text{.}
$
Let $n_{h-1}\left( R\right) -k_{1}=k_{2}$. Construct a sequence $\left\{
x_{i}\right\} _{i=1}^{h-1}$ of elements in
\begin{equation*}
\bigcup _{i=1}^{h}N_{i}\left( R\right) \backslash \left( A\cup B\cup \left(
\bigcup _{i=1}^{h-1}A_{1i}\right) \right)
\end{equation*}
(of length equal to $h-1$) such that $x_{i}\in N_{i}\left( R\right) $ for
all $i=1,...,h-1$. Thus, the sequence is constructed in a way that that no
element of $N_{i}\left( R\right) $ stands next to another element of $%
N_{i}\left( R\right) $, and the last element of the sequence belongs to $%
N_{h-1}\left( R\right) $. Since there are $k_{2}$ sequences of this type,
list these sequences one after the other. By construction, no two
consecutive allocations of this arrangement allocate $j_{1}^{\ast }$ to the
same agent. Join this linear arrangement to the right end of the
arrangement of Step 1. If $n_{h-1}\left( R\right) -k_{1}-k_{2}=n_{1}\left(
R\right) -k_{0}-k_{1}-k_{2}$, then the derived linear arrangement can be
transformed into a circular arrangement by joining its ends. Otherwise, move
to Step 4. For each $i=1,...,h-2$, Let $A_{2i}$ denote the set of elements
of $N_{i}\left( R\right) $ used to construct the sequences. Thus, for each $%
i=1,...,h-2$, $\#A_{2i}=k_{2}$ and $N_{i}\left( R\right) \backslash \left(
A_{1i}\cup A_{2i}\right) $ is the set of allocations that still needs to be
arranged.

\noindent  $\vdots $

\noindent {\bf Step $\ell $:} Then, 
$
n_{1}\left( R\right) -\sum_{i=0}^{\ell -1}k_{i}=n_{2}\left( R\right)
-\sum_{i=1}^{\ell -1}k_{i}\geq ...\geq n_{h-\left( \ell -1\right) }\left(
R\right) -\sum_{i=1}^{\ell -1}k_{i}\text{.}
$
Let $n_{h-\left( \ell -1\right) }\left( R\right) -\sum_{i=1}^{\ell
-1}k_{i}=k_{\ell }$. Construct a sequence $\left\{ x_{i}\right\}
_{i=1}^{h-\left( \ell -1\right) }$ of elements in 
$\bigcup _{i=1}^{h-\left(
\ell -1\right) }N_{i}\left( R\right) \backslash \left( A\cup B\cup \left(
\bigcup _{i=1}^{h-\left( \ell -1\right) }\bigcup _{j=1}^{\ell -1}A_{ji}\right)
\right) $ (of length equal to $h-\left( \ell -1\right) $) such that $%
x_{i}\in N_{i}\left( R\right) $ for all $i=1,...,h-\left( \ell -1\right) $.
Thus, the sequence is constructed in a way that that no element of $%
N_{i}\left( R\right) $ stands next to another element of $N_{i}\left(
R\right) $, and the last element of the sequence belongs to $N_{h-1}\left(
R\right) $. Since there are $k_{\ell }$ sequences of this type, list these
sequences one after the other. By construction, no two consecutive
allocations of this arrangement allocate $j_{1}^{\ast }$ to the same agent.
Join this linear arrangement to the right end of the arrangement of Step $%
\ell -1$. If $n_{h-\left( \ell -1\right) }\left( R\right) -\sum_{i=1}^{\ell
-1}k_{i}=n_{1}\left( R\right) -\sum_{i=0}^{\ell -1}k_{i}$, then the derived
linear arrangement can be transformed into a circular arrangement by joining
its ends. Otherwise, move to Step $\ell +1$. For each $i=1,...,h-\ell $, Let
$A_{\ell i}$ denote the set of elements of $N_{i}\left( R\right) $ used to
construct the sequences. Thus, for each $i=1,...,h-\ell $, $\#A_{\ell
i}=k_{\ell }$ and $N_{i}\left( R\right) \backslash \left( \bigcup _{j=1}^{\ell
}A_{ji}\right) $ is the set of allocations that still needs to be arranged.

\noindent $\vdots $

Since the set of allocations is finite, the above procedure is finite and it
produces a circular arrangement of elements of $F\left( R\right) $ such that
no two consecutive allocations allocate $j_{1}^{\ast }$ to the same agent. $\hfill \square$

\medskip

For each $R\in \mathcal{\bar{R}}$, \Cref{Lem2J*} implies that elements
of $F\left( R\right) $ can be ordered as
$
x\left( 1,R\right) ,...,x\left( m,R\right) \text{,}
$%
with $m=\sum_{i\in N}n_{i}\left( R\right) >1$, such that for all $k=1,...,m$
(mod $m$), if $x_{i}\left( k,R\right) =j_{1}^{\ast }$ for some $i\in N$,
then $x_{i}\left( k+1,R\right) \neq j_{1}^{\ast }$.
Fix any $R^{\prime }\in \mathcal{\bar{R}}$ such that $F\left( R\right) \neq
F\left( R^{\prime }\right) $. We need to consider only the case that $%
\#F\left( R^{\prime }\right) >1$. Suppose that for all $x\left( i,R\right)
\in F\left( R\right) $, there do not exist any agent $\ell $ and any
allocation $z\in \bar{J}$ such that $zP_{\ell }^{\prime }x\left( i,R\right) $
and $x\left( i,R\right) R_{\ell }z$. This implies that for all $x\left(
i,R\right) \in F\left( R\right) $, $L_{\ell }\left( x\left( i,R\right)
,R\right) \subseteq L_{\ell }\left( x\left( i,R\right) ,R^{\prime }\right) $
for all $\ell \in N$. Since $F$ is (Maskin) monotonic, it follows that $%
F\left( R\right) =F\left( R^{\prime }\right) $, which is a contradiction.
Thus, for some $x\left( i,R\right) \in F\left( R\right) $, there exist an
agent $\ell $ and an allocation $z\in \bar{J}$ such that $zP_{\ell }^{\prime
}x\left( i,R\right) $ and $x\left( i,R\right) R_{\ell }z$. Fix any of such $%
x\left( i,R\right) \in F\left( R\right) $. Since by construction of the set $%
\left\{ x\left( 1,R\right) ,...,x\left( m,R\right) \right\} $ we have that
for all $k=1,...m$, with $k\neq i$, it holds that $x\left( k+1,R\right)
P_{j}^{\prime }x\left( k,R\right) $ for some $j$, it follows that $x\left(
i,R\right) $ can be reached via a myopic improvement path at $R^{\prime }$
by any outcome in $x\left( k,R\right) \in \left\{ x\left( 1,R\right)
,...,x\left( m,R\right) \right\} \backslash \left\{ x\left( i,R\right)
\right\} $. Thus, $F$ satisfies {\it rotation monotonicity}.

\end{document}